\documentclass[longauth, printer]{aa}
\pdfoutput=1
\usepackage[varg]{txfonts}
\usepackage{siunitx}
\usepackage{amsmath}
\usepackage{upgreek}
\usepackage{graphicx}
\usepackage{bbold}
\usepackage{color,soul}

\title{Deep Images of the Galactic Center with GRAVITY}

\usepackage{natbib}
\bibpunct{(}{)}{;}{a}{}{,}

\newcommand{\dd}{\mathrm{d}}
\newcommand{\ee}{\mathrm{e}}
\newcommand{\gr}{G\textsuperscript{R}}

\author{GRAVITY Collaboration\thanks{GRAVITY is developed in a collaboration by the Max Planck Institute for extraterrestrial Physics, LESIA of Observatoire de Paris/Universit\'e PSL/CNRS/Sorbonne Universit\'e/Universit\'e de Paris and IPAG of Universit\'e Grenoble Alpes /CNRS, the Max Planck Institute for Astronomy, the University of Cologne, the CENTRA - Centro de Astrofisica e Gravita\c c\~ao, and the European Southern Observatory. \newline
Corresponding authors: J. Stadler (email jstadler@mpe.mpg.de) and A. Drescher (email drescher@mpe.mpg.de).
    }:
R.~Abuter\inst{8}
\and N.~Aimar\inst{2}
\and A.~Amorim\inst{6,12}
\and P.~Arras\inst{17, 26}
\and M.~Baub\"ock\inst{1, 18}
\and J.P.~Berger\inst{5,8}
\and H.~Bonnet\inst{8}
\and W.~Brandner\inst{3}
\and G.~Bourdarot\inst{5,1}
\and V.~Cardoso\inst{12,20}
\and Y.~Cl\'{e}net\inst{2}
\and R.~Davies\inst{1}
\and P.T.~de~Zeeuw\inst{10,1}
\and J.~Dexter\inst{13,1}
\and Y.~Dallilar\inst{1}
\and A.~Drescher\inst{1}
\and F.~Eisenhauer\inst{1}
\and T.~En{\ss}lin\inst{17}
\and N.M.~F\"orster~Schreiber\inst{1} 
\and P.~Garcia\inst{7,12}
\and F.~Gao\inst{1, 19}
\and E.~Gendron\inst{2}
\and R.~Genzel\inst{1,11}
\and S.~Gillessen\inst{1}
\and M.~Habibi\inst{1}
\and X.~Haubois\inst{9}
\and G.~Hei{\ss}el\inst{2}
\and T.~Henning\inst{3}
\and S.~Hippler\inst{3}
\and M.~Horrobin\inst{4}
\and A.~Jim\'enez-Rosales\inst{1,21}
\and L.~Jochum\inst{9}
\and L.~Jocou\inst{5}
\and A.~Kaufer\inst{9}
\and P.~Kervella\inst{2}
\and S.~Lacour\inst{2}
\and V.~Lapeyr\`ere\inst{2}
\and J.-B.~Le~Bouquin\inst{5}
\and P.~L\'ena\inst{2}
\and D.~Lutz\inst{1}
\and F.~Mang\inst{1}
\and M.~Nowak\inst{15,2}
\and T.~Ott\inst{1}
\and T.~Paumard\inst{2}
\and K.~Perraut\inst{5}
\and G.~Perrin\inst{2}
\and O.~Pfuhl\inst{8,1}
\and S.~Rabien\inst{1}
\and J.~Shangguan\inst{1}
\and T.~Shimizu\inst{1}
\and S.~Scheithauer\inst{3}
\and J.~Stadler\inst{1,17}
\and O.~Straub\inst{1}
\and C.~Straubmeier\inst{4}
\and E.~Sturm\inst{1}
\and L.J.~Tacconi\inst{1}
\and K.R.W. Tristram\inst{9}
\and F.~Vincent\inst{2}
\and S.~von~Fellenberg\inst{1}
\and I.~Waisberg\inst{14,1}
\and F.~Widmann\inst{1}
\and E.~Wieprecht\inst{1}
\and E.~Wiezorrek\inst{1} 
\and J.~Woillez\inst{8}
\and S.~Yazici\inst{1,4}
\and A.~Young\inst{1}
\and G.~Zins\inst{9}
}

\institute{
Max Planck Institute for extraterrestrial Physics,
Giessenbachstra{\ss}e~1, 85748 Garching, Germany
\and LESIA, Observatoire de Paris, Universit\'e PSL, CNRS, Sorbonne Universit\'e, Universit\'e de Paris, 5 place Jules Janssen, 92195 Meudon, France
\and Max Planck Institute for Astronomy, K\"onigstuhl 17, 
69117 Heidelberg, Germany
\and $1^{\rm st}$ Institute of Physics, University of Cologne,
Z\"ulpicher Stra{\ss}e 77, 50937 Cologne, Germany
\and Univ. Grenoble Alpes, CNRS, IPAG, 38000 Grenoble, France
\and Universidade de Lisboa - Faculdade de Ci\^encias, Campo Grande,
1749-016 Lisboa, Portugal 
\and Faculdade de Engenharia, Universidade do Porto, rua Dr. Roberto
Frias, 4200-465 Porto, Portugal 
\and European Southern Observatory, Karl-Schwarzschild-Stra{\ss}e 2, 85748
Garching, Germany
\and European Southern Observatory, Casilla 19001, Santiago 19, Chile
\and Sterrewacht Leiden, Leiden University, Postbus 9513, 2300 RA
Leiden, The Netherlands
\and Departments of Physics and Astronomy, Le Conte Hall, University
of California, Berkeley, CA 94720, USA
\and CENTRA - Centro de Astrof\'{\i}sica e
Gravita\c c\~ao, IST, Universidade de Lisboa, 1049-001 Lisboa,
Portugal
\and Department of Astrophysical \& Planetary Sciences, JILA, Duane Physics Bldg., 2000 Colorado Ave, University of Colorado, Boulder, CO 80309, USA
\and Department of Particle Physics \& Astrophysics, Weizmann Institute of Science, Rehovot 76100, Israel
\and Institute of Astronomy, Madingley Road, Cambridge CB3 0HA, UK
\and Department of Physics, Technical University Munich, James-Franck-Straße 1,  85748 Garching, Germany
\and Max Planck Institute for Astrophysics, Karl-Schwarzschild-Str. 1, 85741 Garching, Germany
\and Department of Physics, University of Illinois, 1110 West Green Street, Urbana, IL 61801, USA
\and Hamburger Sternwarte, Universität Hamburg, Gojenbergsweg 112, 21029 Hamburg, Germany
\and CERN, 1 Esplanade des Particules, Genève 23, CH-1211, Switzerland
\and Department of Astrophysics, IMAPP, Radboud University, 6500 GL Nĳmegen, The Netherlands
}

\abstract{Stellar orbits at the Galactic Center provide a very clean probe of the gravitational potential of the supermassive black hole. They can be studied with unique precision, beyond the confusion limit of a single telescope, with the near-infrared interferometer GRAVITY. Imaging is essential to search the field for faint, unknown stars on short orbits which potentially could constrain the black hole spin. Furthermore, it provides the starting point for astrometric fitting to derive highly accurate stellar positions. Here, we present \gr, a new imaging tool specifically designed for Galactic Center observations with GRAVITY. The algorithm is based on a Bayesian interpretation of the imaging problem, formulated in the framework of information field theory and building upon existing works in radio-interferometric imaging. Its application to GRAVITY observations from 2021 yields the deepest images to date of the Galactic Center on scales of a few milliarcseconds. The images reveal the complicated source structure within the central $100\,\mathrm{mas}$ around Sgr~A*, where we detected the stars S29 and S55 and confirm S62 on its trajectory, slowly approaching Sgr~A*. Furthermore, we were able to detect S38, S42, S60, and S63 in a series of exposures for which we offset the fiber from Sgr~A*. We provide an update on the orbits of all aforementioned stars. In addition to these known sources, the images also reveal a faint star moving to the west at a high angular velocity. We cannot find any coincidence with any known source and, thus, we refer to the new star as S300. From the flux ratio with S29, we estimate its K-band magnitude as $m_\mathrm{K}\left(\mathrm{S300}\right)\simeq 19.0 - 19.3$. Images obtained with CLEAN confirm the detection. To assess the sensitivity of our images, we note that fiber damping reduces the apparent magnitude of S300 and the effect increases throughout the year as the star moves away from the field center. Furthermore, we performed a series of source injection tests. Under favorable circumstances, sources well below a magnitude of 20 can be recovered, while $19.7$ is considered the more universal limit for a good data set.}


\begin{document}

\maketitle

\section{Introduction}
\label{sec:introduction}
The Galactic Center (GC) is a unique laboratory for probing general relativity (GR) \citep{2010RvMP...82.3121G}, where stars orbiting Sagittarius A* (Sgr~A*) serve as clean test particles in the gravitational field of a supermassive black hole (SMBH). With near-infrared (near-IR) interferometry, it is not only possible to peer through the dust obscuring the GC, but to also push the angular resolution beyond a single telescope's diffraction limit. Down to an astrometric accuracy of  $\sim65\,\upmu\mathrm{as}$, this technique has been pioneered by the GRAVITY instrument \citep{2017A&A...602A..94G}, which couples the four 8m telescopes at the ESO Very Large Telescope (VLT). At the available telescope separation of $\lesssim \SI{130}{m}$, Sgr~A* and the stars in its vicinity still appear as point sources, but their positions can be determined by a factor of $\sim 20$ more accurately compared to adaptive optics (AO) assisted imaging with a single telescope of similar size. Most importantly, the high angular resolution of GRAVITY allows to overcome the confusion limit of AO imaging.

There are about $50$ stars with known orbits within $1\,\si{arc~second}$ (as) of Sgr~A* \citep{2017ApJ...837...30G}. The most prominent of them, S2, passed its pericenter in 2018. The close monitoring of this event allowed for the detection of the gravitational redshift \citep{2018A&A...615L..15G, 2019Sci...365..664D} and the Schwarzschild precession \citep{2020A&A...636L...5G} in the S2 orbit. In 2019, S2 still was the brightest source next to Sgr~A* within the GRAVITY field of view (FOV) \citep{2021A&A...645A.127G};  in 2021, it moved to a separation that is sufficiently large such that the two sources cannot be detected in a single pointing. Two other stars, however, approach Sgr~A* and go through their pericenters in 2021 -- S29 and S55 \citep{2017ApJ...837...30G}, which we have detected within $50\,\mathrm{mas}$ from Sgr~A*. In comparison with S2, S29 approaches the black hole even more closely, while S55 has a shorter orbital period \citep{2012Sci...338...84M}. We present updated results from their GR orbits in a second paper \citep{mass_distribution}.

The magnitude at which gravitational redshift and Schwarzschild precession affect stellar orbits are on the order of $\beta^2$ \citep[where $\beta=v/c$,][]{2006ApJ...639L..21Z}, while the Lense-Thirring precession due to the black hole spin falls off faster with the distance to the black hole. It is thus not clear whether any of the known stellar orbits allow for the detection of higher-order GR effects \citep{2010PhRvD..81f2002M,2017ApJ...834..198Z}. A faint star at a smaller radius, on the other hand, could provide an opportunity to measure the spin of the SMBH \citep{2018MNRAS.476.3600W}. The expected number of stars suitable for such a measurement has been estimated around unity from extrapolation of the density profile and mass function observed at the GC \citep{2003ApJ...594..812G, 2013ApJ...764..154D, 2018A&A...609A..26G} to small radii and faint stars, respectively \citep{2018MNRAS.476.3600W}.

With the interferometry approach, each baseline between two telescopes represents a point in Fourier space and the correlation of their signal corresponds to the Fourier transform of the image at this coordinate, in accordance with the well-known van Cittert-Zernike theorem \citep{1934Phy.....1..201V,1938Phy.....5..785Z}. Taking advantage of the Earth's rotation over a longer observing sequence and multi-wavelength observations help to fill the so-called $\left(u,v\right)$-plane. If sufficient prior knowledge on the observed flux distribution exists, model fitting is a powerful method to extract the desired information from the data. Without any prior indication of where  a source might be expected, on the other hand, image reconstruction is the technique of choice in the search for faint, as-yet-unknown stars.

Consequently, deep imaging is essential in pushing the exploration of the GC further. Beyond the quest for faint stars, it is also the method of choice for exploring a lesser known field and serves as the starting point for astrometric fitting. The detection of S62, a slowly moving star at the K-band magnitude of $m_\mathrm{K}\left(\mathrm{S62}\right) \simeq 18.9$, in GRAVITY images reconstructed with the radio-interferometry algorithm CLEAN \citep{1974A&AS...15..417H}, clearly demonstrates the power and value of the imaging approach \citep{2021A&A...645A.127G}. CLEAN views the image as a collection of point sources, whose signal it subtracts iteratively from the measured coherent flux until only the noise is left. The question of where to place those point sources and when to stop iterations is guided by the Fourier inversion of the data. Because the $\left(u,v\right)$-space is only sparsely sampled, this ``dirty image" is usually dominated by the inverse Fourier transform of the sampling pattern (the so-called dirty beam) and only the most prominent sources in the field are apparent. After subtracting their signal, fainter sources become recognizable in the residual images. As such, CLEAN depends on the linearity and invertibility of the Fourier transform that relates the image to the data.

Both conditions are not strictly satisfied for near-IR interferometry, where instrumental and observational effects, such as the finite bandwidth size and optical aberrations introduced by the instrument \citep{2021A&A...647A..59G}, complicate the measurement equation. This is similar to what is known as direction-dependent effects (DDEs) in radio interferometric imaging \citep[see e.g.][]{2008A&A...487..419B, 2011A&A...527A.106S}. Furthermore, the requirement for a linear measurement equation impedes the use of closure quantities \citep{1974ApJ...193..293R}, non-linear combinations of the data on different baselines that are insensitive to any telescope-based errors and thus provide more robust measurements. 

Bayesian forward modeling offers an alternative approach to image reconstruction in which possible flux distributions are quantified by the prior and the optimal image is found from exploration of the joint prior and likelihood, that is, the posterior. In this fashion, only the forward implementation of the measurement equation (and possibly its derivative for effective minimization) are needed, such that non-linear and non-invertible terms can be handled straightforwardly. The method, however, comes at the disadvantage of increased computational costs required to perform the posterior search.

Existing tools for optical and near-IR interferometric imaging, such as MIRA \citep{2008SPIE.7013E..1IT} and SQUEZZE \citep{2010SPIE.7734E..2IB}, perform the posterior search either by descent minimization or with Monte Carlo Markov Chains (MCMC). While it may be faster, the former method is limited to convex likelihood and prior formulations. The MCMC method, on the other hand, becomes very inefficient for high-dimensional problems. For imaging, where every pixel is a free parameter to be inferred, this limits the applicability to rather small grids. Furthermore, these are general-purpose codes intended for the application to a range of instruments and, therefore, they do not account for all instrumental effects known to impact GRAVITY data. Finally, imaging the GC is complicated by the variability of Sgr~A* on a timescale of five minutes \citep{2003Natur.425..934G,2004ApJ...601L.159G, 2005ApJ...628..246E, 2006ApJ...640L.163G,2008A&A...492..337E, 2009ApJ...691.1021D, 2011ApJ...728...37D, 2018ApJ...863...15W, 2020A&A...638A...2G, 2020A&A...635A.143G}. Naively combined over a longer period, the data become inconsistent, but $\left(u,v\right)$-sparsity prevents snapshot images over such a short duration. Rather, a prior model beyond those currently available in public codes is required, which can accommodate flux variations of the central source in a otherwise static field of point sources.

In this paper, we present our new imaging code GRAVITY-RESOLVE (\gr)\footnotemark. It builds upon RESOLVE \citep{2021A&A...646A..84A, 2018arXiv180302174A}, a Bayesian algorithm for radio interferometry, but it is tailored to GC observations with GRAVITY in its measurement equation and its prior model. For exploring the posterior distribution, we employed Metric Gaussian Variational Inference \citep[MGVI,][]{2019arXiv190111033K}, an algorithm that aims to provide a trade-off between robustness to complicated posterior shapes and applicability to high-dimensional problems. We present the details of \gr~in Sect.~\ref{sec:method}, describe the GRAVITY data in Sect.~\ref{sec:data} and apply \gr~to it in Sect.~\ref{sec:results}. The images reveal the rich structure of sources around the SMBH and a previously unknown faint star, moving to the west at high angular velocity. In Sect.~\ref{sec:discussion}, we first compare the results of the new algorithm to image reconstruction with CLEAN, before we discuss the properties and implications of the newly detected star. Finally, we present our conclusions in Sect.~\ref{sec:conclusions}. 
\footnotetext{The \gr~source code is available at \url{https://gitlab.mpcdf.mpg.de/gravity/gr_public.git}}

\section{Imaging method}
\label{sec:method}
The backbone of our method is a Bayesian interpretation of the imaging process, in which the posterior probability for a certain image, $I,$ given the data, $d,$ is proportional to
\begin{equation}
\mathcal{P}\left(\left.I\right|d\right) \propto \mathcal{P}\left(\left.d\right|I\right)\,\mathcal{P}\left(I\right)\,,
\label{eq:method-posterior}
\end{equation}
that is, the prior times the likelihood. The proportionality arises because we have omitted the notoriously difficult-to-calculate evidence term $\mathcal{P}\left(d\right)$, which is insensitive to the particular image realization (we come back to this choice in Sect.~\ref{sec:method-inference}). If the posterior distribution is known, we can easily find the most likely image from it or we can compute the expected image and its variance from the first two posterior moments. In practice, however, the posterior is a very high-dimensional scalar function (each image pixel contributes one dimension), and we cannot evaluate its moments analytically. It is then necessary to resort to numerical methods, such as descent minimization to find a (local) maximum or to sampling techniques.

In this section, we introduce all the necessary ingredients to implement the Bayesian inference scheme in practice. We start with the prior (Sect.~\ref{sec:method-prior}), which is formulated as a generative model such that random samples can be drawn from it. Each sample constitutes a possible realization of the image before knowledge of the data. It is then processed by the instrumental response function (Sect.~\ref{sec:method-response}) to arrive at a prediction of the data. Some time-varying instrumental effects, such as coherence loss over a baseline, cannot be described to sufficient accuracy by a deterministic model, and we account for them in a self-calibration approach (Sect.~\ref{sec:method-selfcal}). The agreement between predicted and actual data is then quantified by the likelihood (Sect.~\ref{sec:method-likelihood}). With this, all the components are in place to compute the posterior of a certain sample. Finally, the exploration of the posterior is performed with the MGVI algorithm (Sect.~\ref{sec:method-inference}). 

\subsection{Prior model for the Galactic Center}
\label{sec:method-prior}
A major challenge in imaging the GC is the variability of Sgr~A*. During very bright epochs, it can outshine ambient stars and the signal more strongly resembles the expectation for a single point source. When Sgr~A* is fainter, on the other hand, signatures of the stars in its vicinity become very apparent. Continuous alteration of the flux, between the two extreme cases, prevents the naive combination of exposures over a longer observing period. We account for it by describing Sgr~A* as a time-variable point source that we superimpose on the actual, static image. 

In our prior model, the position of Sgr~A*, $\vec{s}_\mathrm{SgrA}$, follows a Gaussian distribution with user-defined mean and variance. For the flux $I_\mathrm{SgrA}\left(t,p\right)$, we impose a log-normal distribution, and account for the variability by inferring an independent flux value for each exposure $t$. While it would, in principle, be possible to explicitly model the temporal correlation, namely, along the lines of \mbox{\cite{2020arXiv200205218A}}, this further increases the size and complexity of the parameter space, introduces additional hyper-parameters, and makes the sampling more demanding. We therefore leave the exploration of this option to future studies. Furthermore, the emission from Sgr~A* is known to be polarized, with the polarization also varying on short timescales \citep{2007MNRAS.375..764T, 2010A&A...510A...3Z, 2015A&A...576A..20S, 2018A&A...618L..10G, 2020A&A...643A..56G}. We thus also allow for Sgr~A* to have differing, independent fluxes in each of the two polarization states observed by GRAVITY and which we have labeled as $p$.

The actual image, $I_\mathrm{Img}\left(\vec{s}\right),$ then contains the stars in the vicinity of Sgr~A*, where $\vec{s}$ refers to the angular direction relative to the phase center. It must be large enough to cover the full FOV of the GRAVITY fibers, which have a full width half maximum (FWHM) of about $\SI{65}{mas}$. On the other hand, the pixel scale $\delta\alpha$ needs to be sufficiently small to describe the highest spatial frequencies or smallest scales observed by GRAVITY. The latter requirement is the Nyquist–Shannon theorem to avoid spectral aliasing and imposes $\delta\alpha \leq \lambda/\left(2 b_\mathrm{max}\right)$ \citep{2008SPIE.7013E..1IT}. With the longest baseline of $\SI{130}{m}$ and the smallest used spectral channel at $\lambda_\mathrm{min}=2.11\,\upmu\mathrm{m}$, this requires $\delta\alpha \lesssim \SI{1.7}{mas}$. By using a grid with $256^2$ pixels of \SI{0.8}{mas} size, we can meet both requirements simultaneously at manageable computational costs.

To GRAVITY, stars in the GC appear as unresolved point sources. Consequently, we assume that all pixels in the image are statistically independent. We note that this statement applies to the sky prior model and that correlations between nearby pixels introduced by the instrument's finite spatial resolution are described by the response function (Sect.~\ref{sec:method-response}). Furthermore, we expect to see only $\mathcal{O}\left(1\right)$ stars in the image such that the vast majority of pixels will be dark. We describe this situation by imposing an inverse-Gamma prior on the brightness of each pixel:
\begin{equation}
\mathcal{P}\left(I_\mathrm{Img}\right) = \prod_{i}^{N_\mathrm{pix}} \frac{q^\alpha}{\Gamma\left(\alpha\right)} I_\mathrm{Img}(i)^{-\alpha-1} \exp\left[\frac{-q}{I_\mathrm{Img}(i)}\right]\,.
\label{eq:method-faint-sky-prior}
\end{equation}
Here, the index $i$ labels all pixels in the image, $\Gamma$ is the Gamma function, and $q$ and $\alpha$ are two parameters that need to be specified before applying the code. They determine the a priori probability of encountering a bright star (where larger $\alpha$ implies a smaller probability) and allow for the mode of the distribution, $q/(\alpha+1)$, to be set to some background level.

As discussed at length in the following section, the primary observables in optical and near-IR interferometry are the complex visibilities (cf. Eq.~\ref{eq: method-visibilities}), given by the ratio between the coherent flux of a baseline and the flux on each of its two telescopes. Consequently, the data is only sensitive to the relative brightness of all sources. The maximum possible value for the visibility amplitude equals unity and decreases in the presence of a homogeneous background. With this in mind, we set $q=10^{-4}$ and $\alpha=1.5$. In a typical GC observation, the instrumental visibilities are on the order of 0.5, such that this setup implies $\mathcal{O}(1)$ point sources. The probability of encountering a unit point source in the image of faint sources, on the other hand, is $\sim5\%$. We further chose zero as the mean and unity as the variance for the Sgr~A* log-flux prior, whereby the large flexibility of the log-normal distribution can easily accommodate changes by an order of magnitude. 

In many observations, there are bright stars within the FOV around Sgr~A*, whose positions can be adequately predicted from their known orbits. We incorporate them into our model by allowing for additional point sources with a Gaussian prior on their position and a log-normal flux prior. In contrast to Sgr~A*, we assume the flux to be constant in time and the same for both polarization states. In principle, these stars could also be attributed to the image; modeling them explicitly helps to mitigate pixelization errors and improves the convergence.

Finally, the model describes the spectral distribution of Sgr~A* and the stars as two power laws with different spectral indices. These follow a Gaussian prior whose mean we set to $-1$ and $+2$ for Sgr~A* \citep{2010RvMP...82.3121G, 2018ApJ...863...15W} and the stars respectively. As variance, we assume $3$ for both the stars and Sgr~A*. A mathematical description of the sky model is provided in Eq.~(\ref{eq:appx-response-decomposition}), and the priors on each of its components are noted down explicitly in Eq.~(\ref{eq:appx-inference-prior1}) to Eq.~(\ref{eq:appx-inference-prior2}).

\subsection{Instrumental response function}
\label{sec:method-response}

The primary observable of GRAVITY is the complex visibility, measured as the coherent flux over a baseline, $\vec{b,}$ divided by the flux on each telescope forming that baseline. In an idealized setting, it is related to the sky flux distribution, $I\left(\vec{s}, \lambda\right),$ by
\begin{equation}
v\left(\vec{b}/\lambda\right) = \frac{\int_\mathrm{FOV} \mathrm{d}\vec{s}~ I\left(\vec{s}, \lambda\right)\,\mathrm{e}^{-2\uppi \vec{b}\cdot\vec{s}/\lambda}}{\int_\mathrm{FOV} \mathrm{d}\vec{s}~ I\left(\vec{s}, \lambda\right)} \,,
\label{eq: method-visibilities}
\end{equation}
where $\lambda$ is the wavelength. In reality, instrumental effects render the response equation more complicated. We discuss the modifications qualitatively in the following and the generalized expression is given in Eq.~(\ref{eq:appx-response-full}). 

The relevance of the denominator in Eq.~(\ref{eq: method-visibilities}) is most intuitively understood by considering a homogeneous background flux, which averages out in the numerator term for $\vec{b}\neq 0$ but affects the denominator. We therefore implemented the Fourier transform and the normalization term explicitly in the response function.

Equation~(\ref{eq: method-visibilities}) constitutes a simplification in that not all sources are coupled into the fiber equally. GRAVITY uses single mode fibers, which have a Gaussian acceptance profile, to transport the light from the telescope to the beam combiner \citep{2017A&A...602A..94G}. The coupling of light into the instrument is further affected by diffraction \citep{1988ApOpt..27.2334S, 2002A&A...387..366G} and residual tip-tilt jitter from the AO system \citep{2019A&A...625A..48P}. Both effects lead to a wider FOV than implied by the fiber profile alone. Furthermore, there are static optical aberrations along the GRAVITY light path which impact also the phase of the transmitted light \citep{2021A&A...647A..59G}. These aberrations are different for each of the four telescopes.

Taken together, fiber damping and optical aberrations act as a position dependent phase screen which multiplies the image inside the Fourier transform. With the measurement of these so-called phase- and amplitude-maps in \cite{2021A&A...647A..59G}, we are able to provide a full model of the optical aberrations and fiber damping, which we include in the implementation of the response function.

GC observations with GRAVITY are carried out in low spectral-resolution mode and the effect of the spectral bandwidth on the measurement needs to be taken into account. This is known as bandwidth smearing and it leads to a further modification of Eq.~(\ref{eq: method-visibilities}), where the numerator and denominator are independently integrated over the bandpass, namely, $\int \mathrm{d}\lambda P\left(\lambda\right)$. In the most simple case of flat spectral distributions, bandwidth smearing multiplies the visibility amplitudes by a sinc function. For a more realistic scenario in which the source's spectral distribution is modeled as a power law, bandwidth smearing is described by a generalized complex gamma function and, thus, it also affects the visibility phases. Either way, the effect on the measured visibilities increases with the source distance from the image center. 

We account for the finite bandpass size by evaluating the Fourier transform in Eq.~(\ref{eq: method-visibilities}) at multiple values of $\lambda$, spread equally over a top-hat bandpass and taking the average. The complete response equation and details about its efficient implementation are provided in Appendix~\ref{appx:response}.

\subsection{Self-calibration}
\label{sec:method-selfcal}
There are time-variable instrumental effects that our response function does not capture, such as coherence loss on a baseline or telescope-dependent phase errors, for instance, from atmospheric conditions or instrumental systematics. The former leads to a reduction of the measured visibility amplitude, while the latter impacts the visibility phase. If unaccounted for, either of these effects can significantly degrade the imaging solution.

We resolve telescope-dependent phase errors by working with closure phases \citep{1974ApJ...193..293R},
\begin{equation}
\phi_{i,j,k} = \arg\left(v_{i,j}\,v_{j,k}\,v_{k,i}\right)\,,
\end{equation}
which are formed over a triangle of telescopes $i$, $j,$ and $k$, such that telescope-based errors are canceled out. The VLTI consists of four telescopes, implying that six visibility phases and four closure phases can be measured per spectral channel.

GRAVITY carries out a visibility measurement for each baseline. With regard to the amplitudes, we thus adopted a self-calibration approach in which we infer an independent scaling factor, $C\left(t,\,b\right),$ for each exposure and baseline, $b$. This factor joins the list of free model parameters in Appendix~\ref{appx:model}. We impose a Gaussian prior on the scaling factor with unit mean and standard deviation $0.1,  $ as given by Eq.~(\ref{eq:appx-inference-prior2}).

Visibility amplitudes and closure phases are both translation-invariant, that is to say that in the absence of absolute phase information, any global shift of the image will not affect the posterior. Phase- and amplitude maps as well as bandwidth smearing partly break the degeneracy, but these instrumental effects are not sufficiently strong to reliably center the image, particularly if the brightest sources are located close to the image center. Instead, we fix the position of one point source, usually Sgr~A*, which we refer to as as the anchor for our images. 

\subsection{Likelihood}
\label{sec:method-likelihood}
In the limit of sufficiently high signal-to-noise ratios (S/N), that is $\mathrm{S/N} \gtrsim 2-5$, the noise properties of visibility amplitudes and closure phases are well approximated by a Gaussian distribution \citep{2020ApJ...894...31B}. This condition is well satisfied for typical GC observations with GRAVITY. 

However, only three of the four closure triangles available to GRAVITY are statistically independent. In principle, we need to account for this by selecting a reduced, non-redundant closure set whose cross-correlations are captured by a non-diagonal covariance. In practice, and in the limit of equal S/N on all telescopes, equivalent likelihood contours can be obtained from the full closure set without accounting for cross-correlations by up-scaling the error bars with a redundancy factor $N/3$, where $N$ is the number of telescopes \citep{2020ApJ...894...31B}. In this work, we adopt the latter formulation.

Forward modeling approaches are rather sensitive to underestimated error bars. In the case of our imaging code, we observe that overfitting the data can introduce spurious sources in the images. To avoid any potential bias, we applied a conservative $\mathcal{O}(1)$ factor to the error bars in the first imaging iteration. This factor may differ for closures and amplitudes but it is common to all frames and baselines. We then checked the residuals and adjust the error scaling for subsequent runs, aiming for a reduced $\chi^2$ of $0.95 - 1.0$ for visibility amplitudes. To realize the aforementioned redundancy factor, the targeted reduced $\chi^2$ for the closure phases is smaller than the noise expectation from Gaussian statistics and in the range of $0.7 - 0.75$.

\subsection{Inference strategy}
\label{sec:method-inference}

The goal of the inference is to explore the posterior distribution in Eq.~(\ref{eq:method-posterior}) around its maximum over a very high dimensional parameter space. Indeed, for a data set consisting of $N_\mathrm{exp}$ individual exposures and considering $N_\mathrm{PS}$ static point sources in the prior model, the dimensionality is
\begin{align}
d &= 256^2 &\textrm{image of faint sources}\nonumber\\
&+ 2\times N_\mathrm{exp} &\textrm{Sgr~A* light curves}\nonumber \\
&+2 &\textrm{Sgr~A* position} \nonumber\\
&+ 3\times N_\mathrm{PS} &\textrm{point sources position and flux} \nonumber\\
&+ 6 \times N_\mathrm{exp} &\textrm{amplitude self-calibration} \nonumber \\
&+ 2 &\textrm{spectral indices} \label{eq:method-dof} \\
&\sim 7\times 10^4\nonumber\,.
\end{align}
Here,  the factor of two in the light curves arises from the two polarization states observed by GRAVITY and there are six baselines available. Dealing with such high-dimensional distributions is notoriously difficult and computationally expensive. 

The MGVI algorithm \citep{2019arXiv190111033K} used in this study searches for a multivariate Gaussian distribution $\mathcal{G}\left(\left.\vec{\xi}\right| \vec{\bar{\xi}},\, \Xi\right)$ that best approximates the full posterior. Here, $\vec{\xi}$ are standardized coordinates, thus we mapped each degree of freedom to an auxiliary parameter or excitation $\xi$, whose prior is given by a unit Gaussian with zero mean. The inference targets the excitations rather than the physical quantities, however, the two are uniquely related by a mapping (given in Appendix~\ref{appx:model}). As an advantage, this setup makes it fast and easy to draw samples from the prior.

The mean $\vec{\bar{\xi}}$ and covariance $\Xi$ of the posterior approximation are found in an iterative procedure. Starting at some initial position $\vec{\bar{\xi}}_i$ ($i=0$), MGVI uses a generalization of the Fisher metric (cf. Eq.~\ref{appx:inference-fisher-matrix}) to approximate the covariance. This is a $d\times d$ matrix, and by allowing for non-zero off-diagonal elements, MGVI is able to capture cross-correlations between individual model parameters. Since the explicit storage of $d^2$ matrix elements becomes prohibitive for large inference problems, an implicit representation in the form of a numerical operator is used internally.

In the next step, the mean of the approximate distribution $\vec{\bar{\xi}}$ is updated with the aim to increase the overlap between approximate and true posterior $\mathcal{P}\left(\vec{\xi}|\vec{d}\right)$. This overlap is quantified by the Kullback-Leibler divergence (KL), such that updating $\vec{\bar{\xi}}$ amounts to minimizing the KL, that is,\begin{equation}
\vec{\bar{\xi}}_{i+1} = \min_{\vec{\bar{\xi}}} \int \dd\xi~ \mathcal{G}\left(\left.\vec{\xi}\right| \vec{\bar{\xi}}, \Xi_i\right)\, \ln\left[ \frac{\mathcal{G}\left(\left.\vec{\xi}\right| \vec{\bar{\xi}}, \Xi_i\right)}{\mathcal{P}\left(\left.\vec{\xi}\right| \vec{d}\right)}\right]\,.
\label{eq:method-kl}
\end{equation}
The minimum is insensitive to any multiplicative factors applied to the true posterior as long as they are independent of $\vec{\xi}$, thus, the omission of the evidence term in Eq.~(\ref{eq:method-posterior}), in fact, does not affect our analysis. To estimate the expectation value in the KL numerically, MGVI draws samples from the approximate posterior distribution and replaces the integral in Eq.~(\ref{eq:method-kl}) by the sample mean. Once $\vec{\bar{\xi}}$ has been updated, MGVI computes the covariance at the new position and henceforth alternates between the $\Xi$ and $\vec{\bar{\xi}}$ determination. 

After the final iteration, the primary product is a set of samples distributed according to the approximate posterior distribution. These samples can then be used to compute expectation values and their standard deviations, which we report as our main results.

We note that MGVI leaves it to the user to set the number of samples and minimization steps at each iteration as well as the total number of iterations. A good practice is to start with few samples and steps, then increase both quantities as the inference approaches the posterior maximum. Details about the scheme used for this work are given in Appendix~\ref{appx:hyperparameters}, where we also specify the minimizers and iteration controllers used. For the initial position from which to start the inference, we align all parameters with the mode of their respective priors.

Two obstacles in the minimization procedure demand special attention, namely: the convergence of our results and the possibility of multi-modal posterior distributions. In case of the latter, the Gaussian approximation obtained from MGVI captures one typical posterior mode. Exploring multi-modal distributions in such high-dimensions is genuinely a very difficult, computationally expensive problem for which there currently exists no standard practice. To maintain some handle on the multi-modality of the posterior and to judge how well the algorithm has converged, we exploit the inherent stochasticity of MGVI that arises when the KL is estimated from random samples. As our results below indicate, changing the random seed from which samples are drawn can nudge the algorithm to explore different posterior modes. In addition, poorly converged runs -- which are typically more noisy and might contain over-fitted sources or might fail to detect a faint source -- also occur just for individual random seeds, such that they can be judged and eliminated by comparing results from multiple random seeds.

For a typical data set, we performed ten independent imaging runs with different random seeds. This number is large enough to capture the dominant modes of the posterior but too low to determine their relative weights reliably. Instead, we used the fact that we have multiple data sets available and judged the images by their consistency over the full 2021 observing period.

\begin{figure*}
        \centering
        \includegraphics[]{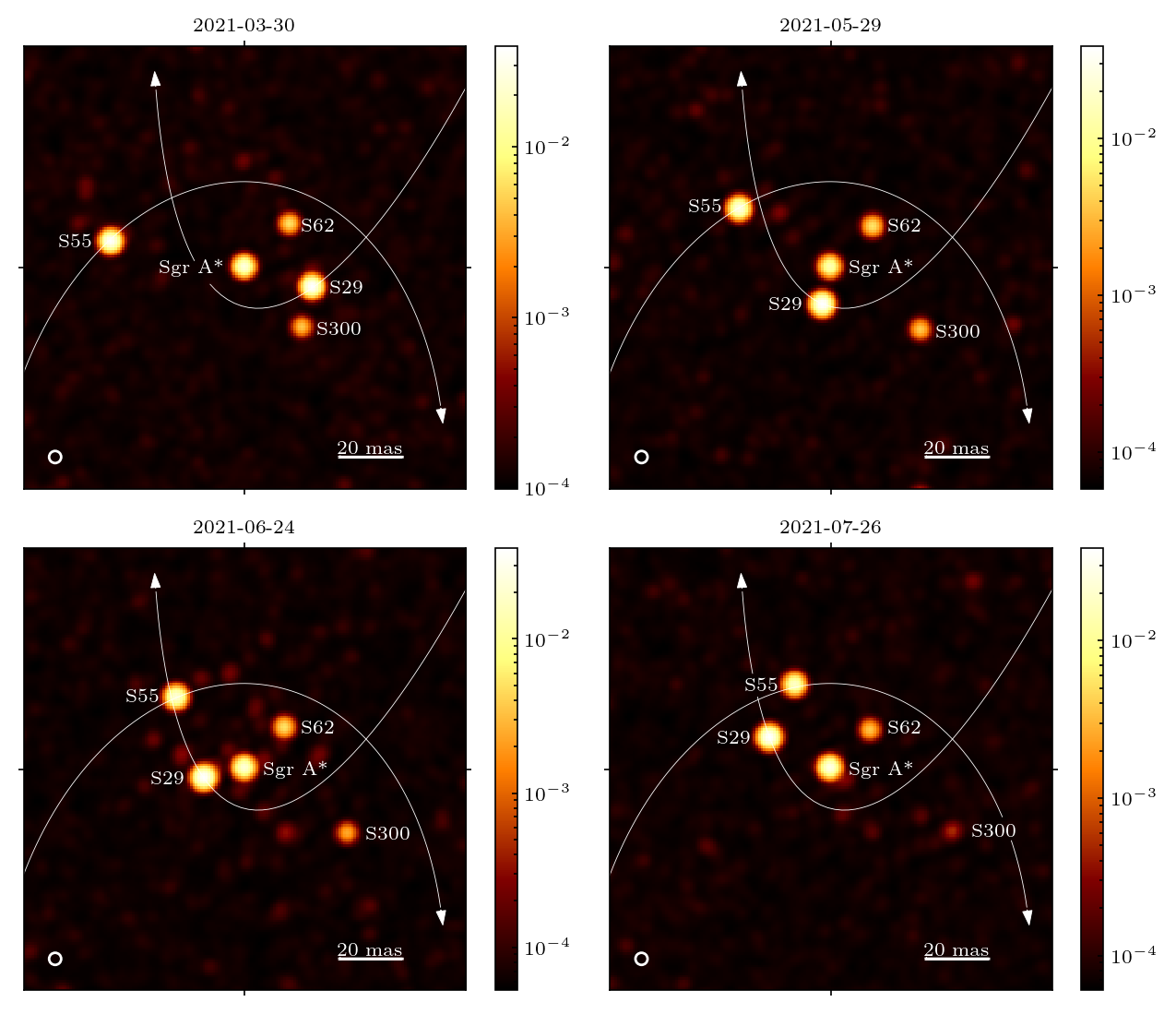}
        \caption{Images for one night per observing epoch in 2021, where north is up, east to the left, and the flux normalized to S29. For each night, we performed ten imaging runs with varying random seeds and then picked the cleanest result with the highest degree of consistency over the full year. Images are computed as mean over all samples in the chosen run, centered on Sgr~A*, and smoothed with a Gaussian of $1.6\,\mathrm{mas}$ standard deviation, whose FWHM is indicated in the bottom left corner. Sgr~A* varies in flux, and we here show its mean brightness over all exposures and polarization states. On each image, we overdraw the orbits of S29 and S55 and the labels of all identified sources to give a better orientation. We note that due to the large dynamical range and the logarithmic color scaling, sources appear more widespread.}
        \label{fig:results-images}
\end{figure*}

\section{Data}
\label{sec:data}
For this study, we considered GC observations with GRAVITY from 2021. They are separated into four epochs taking place at the end of March, May, June, and July, respectively. The data were obtained with GRAVITY's low spectral-resolution, split-polarization mode. Each exposure consists of 32 individual frames with $\SI{10}{\s}$ integration time. Most of them are centered on Sgr~A*, and the Sgr~A* observing blocks are bracketed by S2 and R2 pointings that are required for calibration and to monitor the S2 orbit. The data were reduced by the GRAVITY standard pipeline and calibrated with a single, carefully chosen S2 exposure from the same night.

We used the data to obtain night-wise images of Sgr~A* and its vicinity. To have sufficient $\left(u,v\right)$-coverage, we required $\mathcal{O}(10)$ exposures at least and selected those nights with a larger number of Sgr~A* pointings available. The selection is summarized in Table~\ref{tab:data-observations-sgra}.

\begin{table}
\caption{GC observing nights selected for Sgr~A* imaging.}
\label{tab:data-observations-sgra}
\centering
\begin{tabular}{c c c c c}
Date & \shortstack{Available\\ exposures} & $N_\mathrm{exp}$ & $\left\langle\sigma_\rho\right\rangle$ & $\left\langle\sigma_\phi\right\rangle$\\
\hline
2021-03-30 & 13 & 13 & $4.5\times10^{-2}$ & $33.2^\circ$\\
2021-05-22 & 20 & 19 & $3.3\times10^{-2}$ & $10.5^\circ$\\
2021-05-29 & 22 & 21 & $3.5\times10^{-2}$ & $19.3^\circ$\\
2021-05-30 & 20 & 20 & $3.7\times10^{-2}$ & $24.3^\circ$\\
2021-06-24 & 32 & 29 & $3.6\times10^{-2}$ & $20.7^\circ$\\
2021-07-25 & 28 & 22 & $3.1\times10^{-2}$ & $15.3^\circ$\\
2021-07-26 & 20 & 20 & $3.6\times10^{-2}$ & $19.6^\circ$\\
2021-07-27 & 20 & 19 & $2.8\times10^{-2}$ & $10.3^\circ$
\end{tabular}
\tablefoot{Each exposure amounts to a \SI{320}{\s} integration time. For some nights, we deselected exposures which were affected by bad weather or instrumental problems, such that the number of exposures used in the imaging, $N_\mathrm{exp}$, is smaller than the total number of exposures available. The last two columns give the mean standard deviation of the closure phases and amplitudes. If a scaling factor was applied to the error bars, this is included.}
\end{table}

\begin{table*}
\caption{Overview of the mosaicing data set.}
\label{tab:data-observations-widefield}
\centering
\begin{tabular}{c c c c c c}
Name & Date & \# of exposures & \shortstack{Pointing offset w.r.t\\ Sgr~A* (RA, Dec [mas])} & preset sources & anchor\\
\hline
S2 & 2021-05-29 \& $30$ & $13$ & $24.8$, $142.4$ & S2 & S2 \\
NW & 2021-07-29 & $8$ & $-45.0$, $45.0$ & Sgr~A*, S29, S55, S42 & S42 \\
SE & 2021-07-29 & $8$ & $45.0$, $-45.0$ & Sgr~A*, S29, S55 & Sgr~A*\\
mid & 2021-07-29 & $7$ & $-24.8$, $-31.4$ & Sgr~A*, S29, S55 & Sgr~A*\\
S38 & 2021-07-25 \& $26$ & $8$ & $-38.6$, $-76.8$ & S38 & S38
\end{tabular}
\tablefoot{The second to last column lists all sources which are modeled as point sources with a Gaussian position prior. To break the translation invariance inherent to closure phases and visibility amplitudes, we fix the location of one bright point source, which we call the anchor for that image.}
\end{table*}

The Sgr~A*-centered images from March to June revealed a faint object, moving to the west at high angular velocity. To track this new star and to also scan a wider field, we performed a series of pointings in July, where we offset the fiber from Sgr~A*. Further, to the south west of Sgr~A* there is S38 with a separation too large to be detected in the Sgr~A*-centered images. For the July observing run, there are a sufficiently large number of S38-centered exposures available to attempt reconstructing an image from them. Just as with the Sgr~A*-centered exposures, the data were reduced by the standard pipeline and calibrated with a single S2 file. Finally, we also considered S2-pointings from May, which were calibrated with R2. The so-called ''mosaicing data set" is summarized in Table~\ref{tab:data-observations-widefield}.

\section{Results}
\label{sec:results}

\begin{figure*}
\sidecaption
\includegraphics[]{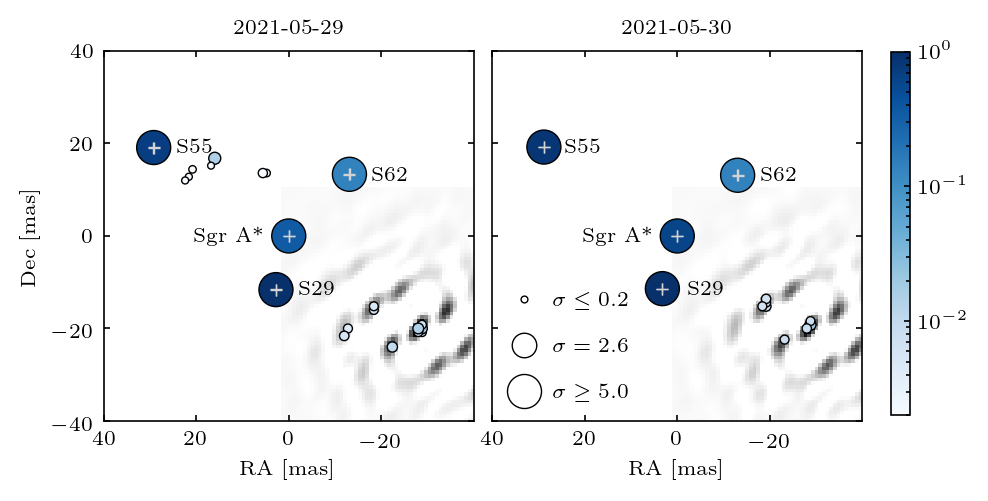}
\caption{Combined view on all imaging runs from a single night (see Appendix~\ref{appx:probabilistic-results} for details on the illustration). The symbol color indicates the flux of a source candidate, normalized to S29, while the symbol size represents its significance. The Sgr A* flux depicted here has to be multiplied by the light curve at each exposure to arrive at the true flux ratio. For stars that are modeled as a point source, the position uncertainty is indicated in gray. Sources are shown superimposed on the dirty beam pattern of the imaging night, which we have centered at the position of S300.}
\label{fig:res-multimodality}
\end{figure*}

We apply the \gr~imaging algorithm (introduced in Sect.~\ref{sec:method}) to obtain night-wise images from the data summarized in Tables~\ref{tab:data-observations-sgra} and~\ref{tab:data-observations-widefield}. Here, we first focus on observations with Sgr~A* at the center and come back to the mosaicing data set later on.

The final output of the MGVI algorithm is a set of samples drawn from the approximate posterior. Our best image is computed as the mean over these samples and we add the flux of all additional point sources to the appropriate pixels. Because of the normalization term in the instrumental response function (cf. Eq.~\ref{eq: method-visibilities}), the likelihood is insensitive to a global scaling of the flux. In a post-processing step, we therefore normalize our images to the flux of S29 \citep[$m_\mathrm{K}\left(\mathrm{S29}\right) \simeq 16.6$, ][]{2021A&A...645A.127G}, which is the brightest static source in the field. The implementation of the response function presented in Appendix~\ref{appx:response} explicitly takes into account fiber damping, which reduces the flux transmission for off-center sources; thereby, our images are automatically corrected for this effect. Finally, we smoothed the images with a Gaussian of $\SI{1.6}{mas}$ standard deviation. This corresponds to the typical size of the CLEAN beam, that is, a Gaussian fit to the central part of the dirty beam pattern.

\subsection{Detection of S29 and S55 within \SI{50}{mas} of Sgr~A*}
In 2021, there are two relatively bright stars in the FOV around Sgr~A*, which are S29 with a K-band magnitude of $m_\mathrm{K}\left(\mathrm{S29}\right) \simeq 16.6$ \citep{2021A&A...645A.127G} and S55 with $m_\mathrm{K}\left(\mathrm{S55}\right) \simeq 17.6$ \citep{2017ApJ...837...30G}. In particular the cross-identification of S29 in earlier AO-based NACO images has been subject of debate recently \citep{2021ApJ...918...25P}. We discuss this in Appendix~\ref{sec:appx-identification}.

The orbital motions of S29 and S55 around the black hole are overdrawn on the images in Fig.~\ref{fig:results-images}. In a couple of test runs, we find that our algorithm is easily capable of detecting each of them. The detection of S29 and S55 in the 2021 Sgr~A*-centered exposures also is a prerequisite and starting point for astrometric fitting to extract high-accuracy stellar positions.

In what follows, we model S29 and S55 as point sources, superimposed on the image as described in Sect.~\ref{sec:method-prior}. Each of the sources has a Gaussian prior on its position, whose mean we obtain either from orbit predictions or from the pixel position in a pre-imaging run. To avoid any bias arising from the latter procedure, we chose a deliberately large standard deviation for the sources' position priors of $\SI{2.4}{mas}$ in RA and Dec. This corresponds to three pixels in our image and is larger than the beam width in any of the observations considered. Since Sgr~A* is fixed at the image center to break the translation invariance inherent to closure phases, mapping the stars' separation vector to an image position is straightforward.

\subsection{Detection of S62}
\label{sec:results-s62}
In \cite{2021A&A...645A.127G}, CLEAN images revealed a 18.9 K-band magnitude star which slowly approaches Sgr~A* and was identified as S62. Extrapolating the motion observed in 2019, we expect S62 at $\mathrm{RA} = \left(-14.1\pm0.4\right)\si{mas}$, $\mathrm{Dec} = \left(13.6\pm0.8\right)\si{mas}$ in March 2021. By July, it should move to $\mathrm{RA}=(-13.2\pm0.4)\,\si{mas}$, $\mathrm{Dec}=\left(12.3\pm0.9\right)\si{mas}$.

An important test for the new imaging method is whether it is also able to detect S62. Indeed, for all nights we infer flux at the expected position. This detection is very robust, namely, $> 5\sigma$ for almost all random seeds and even holds if the error bars are moderately over-scaled. To determine the S62 coordinates beyond the accuracy of the pixel size, we then include it into the set of point sources inferred on top of the image and perform ten imaging runs with varying random seeds for each night. The results are consistent between all runs of an individual night. We are thus able to combine the samples from all ten runs into an estimate on the mean S62 position and its variance. This is summarized in Table~\ref{tab:results-s62} and it does match  the prediction from \cite{2021A&A...645A.127G} very well.

Furthermore, we use the positions in Tab~\ref{tab:results-s62} to provide an updated on the motion of S62. From a linear fit that also considers the results from 2019 observations \citep{2021A&A...645A.127G}, we obtained the following for the relative velocity w.r.t. Sgr~A*:
\begin{align*}
&v_\mathrm{RA} = \left(2.97 \pm 0.05\right)\,\si{mas/yr}\,,\\
&v_\mathrm{Dec} = \left(-3.58 \pm 0.09\right)\,\si{mas/yr}\,.
\end{align*}

The slow linear motion of S62, observed with GRAVITY consistently in 2019 and 2021, does not fit a star with a 9.9 year orbital period as reported in  \mbox{\cite{2020ApJ...889...61P}}. In Appendix~{\ref{sec:appx-identification}}, we explain in detail the cross-identification of all further sources in the FOV, S29, and S55, namely, between GRAVITY and earlier AO-based images. Thus, the GRAVITY images do not support the existence of a star that orbits Sgr~A* with a 9.9 year period.

\begin{table}
\caption{Separation between S62 and Sgr~A* obtained from imaging runs in which S62 was modeled as point source with a Gaussian position prior.}
\label{tab:results-s62}
\centering
\begin{tabular}{c c c}
Epoch & RA $\left[\si{mas}\right]$ & Dec $\left[\si{mas}\right]$\\
\hline
2021.2453 & $-13.85\pm0.11$ & $14.00\pm0.10$\\
2021.3902 & $-13.03\pm0.05$   & $13.12\pm0.06$\\
2021.4093 & $-13.12\pm0.06$   & $13.33\pm0.07$\\
2021.4120 & $-13.05\pm0.06$   & $13.13\pm0.08$\\
2021.4803 & $-12.78\pm0.04$  & $12.84\pm0.05$\\
2021.5649 & $-12.60\pm0.04$  & $12.67\pm0.05$\\
2021.5657 & $-12.74\pm0.05$  & $12.64\pm0.07$\\
2021.5703 & $-12.71\pm0.09$  & $12.57\pm 0.14$
\end{tabular}
\tablefoot{The epoch is computed as mean over all exposures used for the image. For each night we performed ten \gr~runs with varying random seeds and combined the samples from all runs into an estimate of the S62 position and its standard deviation. We note that the standard deviation only accounts for the statistical position uncertainty, but not for any systematic error.}
\end{table}

\subsection{Discovery of S300, a faint fast-moving star}
\label{sec:results-s300-detection}

\begin{figure}
\centering
\includegraphics[]{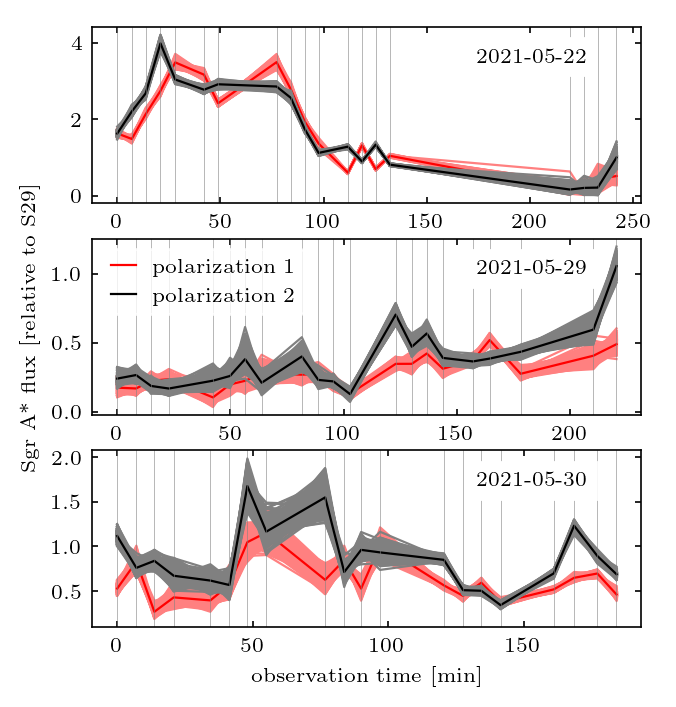}
\caption{Light curves inferred with the \gr-code for the May observing nights. Light lines in gray and red indicate the individual samples and dark lines show the sample mean. Here, we combined all samples from all imaging runs of a particular night. The time of the individual exposures is indicated by vertical lines. We note that S2 is about $11$ times as bright as S29.}
\label{fig:results-lcmay}
\end{figure}
The main goal of our imaging analysis is to search the vicinity of Sgr~A* for faint, as-yet-unknown stars. To this end, we used the same ten imaging runs per night from which we measured the S62 position (detailed in Sect.~\ref{sec:results-s62}). A representative image for each epoch is shown in Fig.~\ref{fig:results-images}. In Appendix~\ref{appx:probabilistic-results}, we provide a complementary view of the imaging results for all nights listed in Table~\ref{tab:data-observations-sgra}, which accounts for the statistical nature of our algorithm. 

In addition to the four expected sources -- Sgr~A*, S29, S55, and S62 -- our images contain a fifth object that is fainter than all aforementioned stars. It moves to the west with a high angular velocity, so that the change in position can be recognized very clearly between individual months. In the following, we discuss this detection for each epoch separately.

\subsubsection{March 2021} 
Due to the limited observability of the GC in March, this is the data set with the sparsest $\left(u,v\right)$-coverage. However, it is also the epoch where the new source is closest to the center of the FOV and thus it is the least affected by fiber damping. Of the ten imaging runs we performed, five detected the new source as a single bright pixel with high significance ($> 5\sigma$). Its location on the grid can vary by one pixel between runs. Two further runs infer flux at the same position, but smeared out over multiple neighboring pixels.

\subsubsection{May 2021}
In Sect.~\ref{sec:method-inference}, we mention the possibility of multi-modal posterior distributions. Indeed, the only data set where we have clear signs of such an issue are the three Sgr~A*-centered images from May 2021. 

All ten imaging runs for May 29 infer a single bright pixel to the south west of Sgr~A* at a high significance. In three instances, however, the location of this pixel is shifted towards the image center. A similar situation arises for May 30. In four instances, the location of the bright pixel coincides with the position found for the previous night, otherwise it is shifted inwards. 

We illustrate this situation in Fig.~\ref{fig:res-multimodality}, where we have combined the samples from all ten imaging runs into a single figure for each night. Even though the new source is detected with a high significance in individual \gr-runs, the fact that its location varies between runs makes the overall significance estimate decrease in Fig.~\ref{fig:res-multimodality} (cf. Appendix~\ref{appx:probabilistic-results}). The sources detected in the image are superimposed on the dirty beam pattern of the respective nights to illustrate the reason behind the observed multi-modality. This shows that the inward-shifted detections of the new source correspond to side-maxima or side-lobes of the dirty beam pattern. 

On May 29, the algorithm apparently is somewhat more successful at disentangling the true source position and its side-lobes than on May 30. There are two possible reasons for this, the first being that the data set for May 29 contains one more exposure and on average exhibits smaller error bars than that of May 30, as Table~\ref{tab:data-observations-sgra} indicates. Secondly, the light curves that we obtain as part of the inference are shown in Fig.~\ref{fig:results-lcmay} for all May nights. They disclose that Sgr~A* was slightly brighter on May 30 than on the 29th. When the complex visibilities are dominated by the central source more strongly, the detection of the faint source further out in the field becomes more difficult.

The images that we infer for May 22 also are consistent with that line of reasoning. As Fig.~\ref{fig:results-lcmay} shows, Sgr~A* went through a moderate flare in the beginning of the night. While all imaging runs for May 22 clearly exhibit some flux to the south west of Sgr~A*, no consistent source position can be identified from the comparison of multiple runs (cf. Fig.~\ref{fig:appx-results-summary}).

\begin{figure}
        \centering
        \includegraphics[]{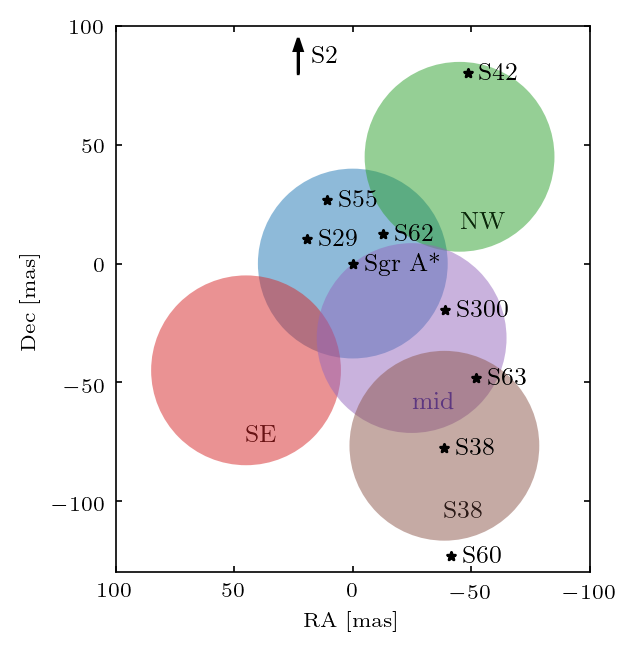}
        \caption{Summary of pointings in the mosaicing data set in the relation to the positions of all detectable stars during the July observing run. For completeness, we further include the Sgr~A*-centered exposures (blue). Colored circles indicate the individual pointings and have diameters of \SI{40}{mas}, which is half the extent of the images in Fig.~\ref{fig:results-offpointing-images}.}
        \label{fig:res-offpointings-summary}
\end{figure}

\begin{figure*}
\centering
\includegraphics[]{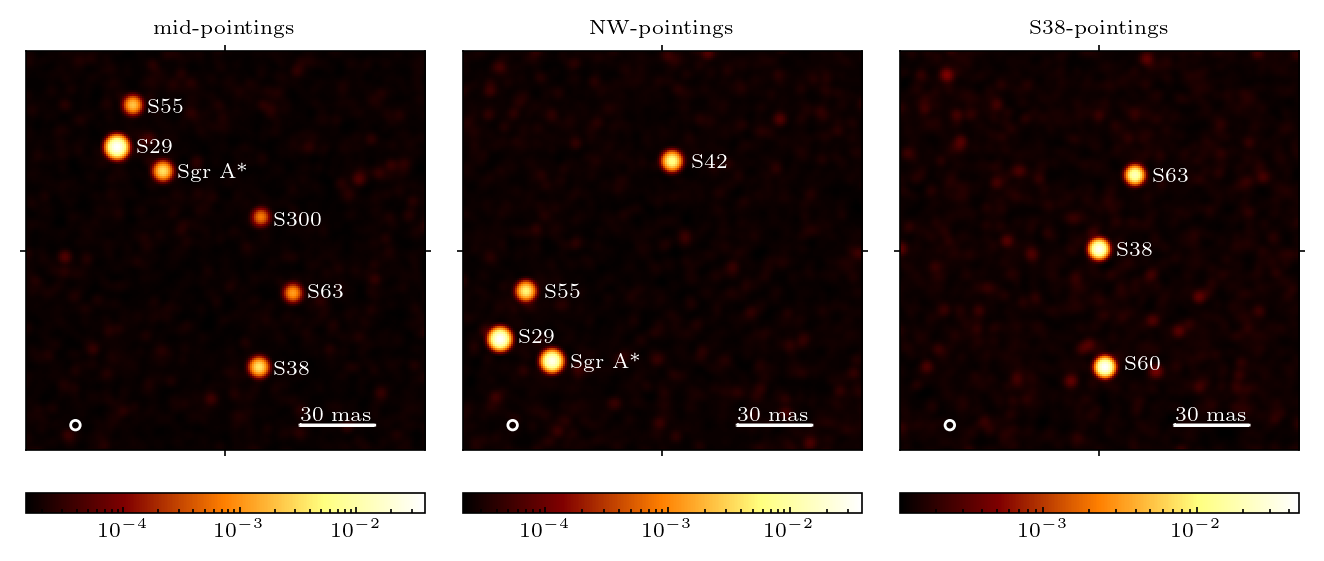}
\caption{Images obtained for the mosaicing data set, where north is up and east to the left. The flux is normalized by S29 in the mid and northwest pointings and by S38 in the rightmost panel. Images are computed as sample mean over all samples of a single \gr~run and have been convolved with a Gaussian of $1.6\,\mathrm{mas}$ standard deviation whose FWHM is indicated in the bottom left corner. The mid- and the S38-pointings extend the Sgr~A*-centered images to the south west. S62 is not detectable in these images due to a combination of two effects. First, at this large distance from the image center, the already faint flux is further damped by the fiber profile; and secondly, the overall sensitivity of the images is reduced in comparison to the Sgr A* pointings because of the smaller number of exposures available.}
\label{fig:results-offpointing-images}
\end{figure*}

\subsubsection{June 2021}
For the large June imaging data set, all ten imaging runs exhibit flux at the same location to the south west of Sgr~A*. In five instances, this flux is detected in a single pixel with high significance ($>5\sigma$). Its position scatters by at most a pixel. In the other five instances, the flux is spread out over several connected pixels, which exhibit larger flux variations between the individual samples.

Since the source has moved further away from the field center and is more strongly affected by fiber damping than it was in May, the question arises why no similar multi-modality of the posterior is observed. On June 24, we have a much better $\left(u,v\right)$-coverage and the dirty beam pattern becomes considerably smoother without the isolated, strongly peaked side-maxima that induce the misplacement in the May images.

\subsubsection{July 2021}
We are able to detect the new source only in one of the July observing nights which is July 26. Even so, it is found in five of the ten imaging runs at a low significance smeared out over multiple pixels. Correspondingly, the source appears fainter in the final image of Fig.~\ref{fig:results-images}. Since the new star is most strongly affected by fiber damping in July, this loss of sensitivity does not come as a surprise. We discuss it further in Sect.~\ref{sec:results-injection}.

\begin{table}
\caption{Position of S300, the newly detected star, with respect to Sgr~A*.}
\label{tab:results-s300-pos-flux}
\centering
\begin{tabular}{c c c c}
Epoch & RA [mas] & Dec [mas] & flux/S29\\
\hline
2021.2453 & $-18.0\pm0.8$ & $-19.6\pm0.8$ & $0.11\pm0.01$\\
2021.4093 & $-28.4\pm0.8$ & $-20.4\pm0.8$ & $0.09\pm0.01$\\
2021.4120 & $-28.4\pm0.8$ & $-18.8\pm0.8$ & $0.08\pm0.01$\\
2021.4803 & $-33.4\pm0.8$ & $-20.4\pm0.8$ & $0.05\pm0.02$\\
2021.5657 & $-39.6\pm0.8$ & $-19.6\pm0.8$ & --
\end{tabular}
\tablefoot{The flux ratio relative to S29, is already corrected for fiber damping and the epoch is computed as mean over all exposures used in the imaging.}
\end{table}

We summarize the source position inferred for each night in Table~\ref{tab:results-s300-pos-flux}, along with its flux relative to S29. For the latter, we selected all runs in which the new source is found as a single bright pixel in the correct location and combined all samples in these runs. That is, we exclude runs from the flux estimate that failed to identify the new source, where the new source was misplaced at a side-lobe, or where its flux was smeared out over multiple pixels. 

The position and motion of the new star matches none of the known S-stars \citep{2017ApJ...837...30G} and we conclude that we have detected a new star, which we refer to as S300. We further discuss the possibilities for its nature in Sect.~\ref{sec:discussion-s300}.

\subsection{Images for the mosaicing data set}
\label{sec:results-mosicing}
To produce images for the mosaicing data set, we used the exact same strategy as for the Sgr~A*-centered exposures. The pointing directions and the known stars in the field are summarized in Fig~\ref{fig:res-offpointings-summary}. In Table~\ref{tab:data-observations-widefield}, we list all stars that we model as point sources with a Gaussian position prior and the anchor, whose positional variance we set to zero in order to break the translation invariance inherent to closure phases and visibility amplitudes (cf. Sect.~\ref{sec:method-selfcal}).

\begin{table*}
\caption{Orbital elements for six stars detected in the 2021 images.}
\label{tab:results-orbital-elements}
\centering
\begin{tabular}{lcccc}
Star& $a$[mas]& $e$& $i\,[^\circ]$& $\Omega\,[^\circ]$\\
\hline
S2&$124.95\pm0.04$&$0.88441\pm0.00006$&$134.70\pm0.03$&$228.19\pm0.03$\\
S29&$397.50\pm1.56$&$0.96927\pm0.00011$&$144.37\pm0.07$&$7.00\pm0.33$\\
S38&$142.54\pm0.04$&$0.81451\pm0.00015$&$166.65\pm0.40$&$109.45\pm1.00$\\
S42&$411.42\pm7.14$&$0.77385\pm0.00309$&$39.57\pm0.19$&$309.60\pm1.06$\\
S55&$104.40\pm0.05$&$0.72669\pm0.00020$&$158.52\pm0.22$&$314.94\pm1.14$\\
S60&$489.21\pm18.55$&$0.77733\pm0.00806$&$126.60\pm0.15$&$178.03\pm0.80$\\
\newline
&  $\omega\,[^\circ]$& $t_P$[yr]& $T$[yr] & $m_K$\\
\hline
S2 & $66.25\pm0.03$&$2018.38\pm0.00$& $16.046\pm0.001$ &13.95\\
S29&$205.79\pm0.33$&$2021.41\pm0.00$& $91.04\pm0.54$ &16.6\\
S38&$27.17\pm1.02$&$2003.15\pm0.01$&  $19.55\pm0.01$ &17.\\
S42&$48.29\pm0.46$&$2022.12\pm0.02$&  $95.9\pm2.5$&17.5\\
S55&$322.78\pm1.13$&$2009.44\pm0.01$& $12.25\pm0.01$ &17.5\\
S60&$30.43\pm0.21$&$2023.62\pm0.05$&  $124.3\pm7.1$ &16.3\\
\end{tabular}
\tablefoot{The parameters listed are the semi major axis, eccentricity, inclination, position angle of the ascending node, longitude of periastron, epoch of periastron passage, orbital period, and K-band magnitude. The stars S62 and S300 are not listed here, as their movement observed so far is consistent with a linear motion. Images serve as starting point for high-accuracy astrometric fitting, and the orbits are determined from positions provided by the latter. }
\end{table*}

In contrast to the Sgr~A*-centered images and with exception of the S2 pointings, we now have fewer than ten exposures available for each individual image. We thus expect (and also go on to find) that the solutions become more noisy and that there is a greater variability between the ten runs which we perform for each data set. The images obtained for the mid, NW, and S38 pointings are shown in Fig.~\ref{fig:results-offpointing-images}, and the full statistical view over all results is provided in Appendix~\ref{appx:probabilistic-results} and in Fig.~\ref{fig:appx-results-offpointings-summary}.

The mid pointings were specifically designed to provide an additional test of the S300 detection. The fiber position is such that S300 is close to the center of the field, but Sgr~A*, S63, and S38 can also be observed within the same pointing; Fig.~\ref{fig:results-offpointing-images} very clearly shows all the expected sources. The new source, S300, is detected with high significance ($> 5\sigma$) in each of the imaging runs, its position varies by at most one pixel in RA and two pixels in Dec. Combining all mid-pointing imaging runs, we obtain:
\begin{align*}
&\mathrm{RA} = \left(-38.8 \pm 0.8\right)\si{mas}\,, \\
&\mathrm{Dec} = \left(-19.4 \pm 1.6\right)\si{mas}\,,\\
&\mathrm{flux/S38} = 0.15 \pm 0.10\,, 
\end{align*}
which is fully consistent with the position obtained from Sgr~A*-centered images in Table~\ref{tab:results-s300-pos-flux}.

An important difference to the Sgr~A*-centered images, in particular, for the mid and S38 pointings, is that we did not pre-set all known sources in the field. In the former case, it was left to our algorithm to detect S300, S63, and S38; in the latter case, S63 and S60. In this context, the images in Fig.~\ref{fig:results-offpointing-images} also demonstrate how \gr~ is able to orient in a field with limited prior knowledge about the overall source structure. Apart from the stars that are sketched in Fig.~\ref{fig:res-offpointings-summary} -- namely Sgr~A*, S2, S29, S38, S42, S55, S60, S63, and S300 -- we do not detect any other objects.

The detection of all the aforementioned sources in the images also serves  as a starting point for astrometric fitting to obtain high accuracy source positions \citep[see][for details on the fitting procedure]{2020A&A...636L...5G}, which, in turn, allows us to determine the stars' orbits. We give an update on the orbital elements of all stars detected in the 2021 images in Table~\ref{tab:results-orbital-elements}.

\subsection{Sensitivity estimation}
\label{sec:results-injection}

As a first step to estimate the sensitivity in our images, we performed a series of injection tests that are reported in Appendix~\ref{appx:injection-test}. They consider sources at two different locations and four magnitudes, between $19.7$ and $22.7$, which are inserted into the May 29 data set. 

The ability of \gr~to recover the injected source depends on its position. In the first scenario, the new star is located close to S300, and we managed a high-significance detection only at $19.7$th magnitude. In addition, we observed a larger scatter around S300, and more flux is placed in its side-lobes. In the second case, the source is injected at same distance but to the north east of Sgr A*. Here, we can reach significantly deeper and manage a robust detection even at $21.0$th magnitude.

We can also use S300 itself to estimate the sensitivity in our images. As the star moves away from the field center, it is more strongly affected by fiber damping and appears fainter to the GRAVITY instrument. Since S300 is most robustly detected close to the field center, we use the images from March, May, and the mid pointing in July to estimate its magnitude from which we obtain $m_\mathrm{K}\left(S300\right) \simeq 19.0 - 19.3$. The corresponding apparent magnitudes, that is, the magnitude corrected for fiber damping, are listed in Table~\ref{tab:results-s300-apparent} for all observing epochs.

\begin{table}
\caption{Apparent magnitude of S300 in all four 2021 observing epochs.}
\label{tab:results-s300-apparent}
\centering
\begin{tabular}{c c c}
Date & min $m_\mathrm{K,app}$ & max $m_\mathrm{K,app}$\\
\hline
March 2021 & 19.3 & 19.7\\
May 2021 & 19.6 & 19.9\\
June 2021 & 19.8 & 20.1\\
July 2021 & 20.1 & 20.5
\end{tabular}
\tablefoot{The apparent magnitute of a star accounts for corrections due to fiber damping if it is not located at the center of the FOV. The minimum and maximum values correspond to our bracketing estimates for the intrinsic S300 brightness.}
\end{table}

Until June 2021, S300 is very robustly detected in our images which also matches the results of the mock tests above. In July, on the other hand, the apparent magnitude of S300 is already below $20$, and we only managed a weak detection at low significance.

\begin{figure*}
        \centering
        \includegraphics[]{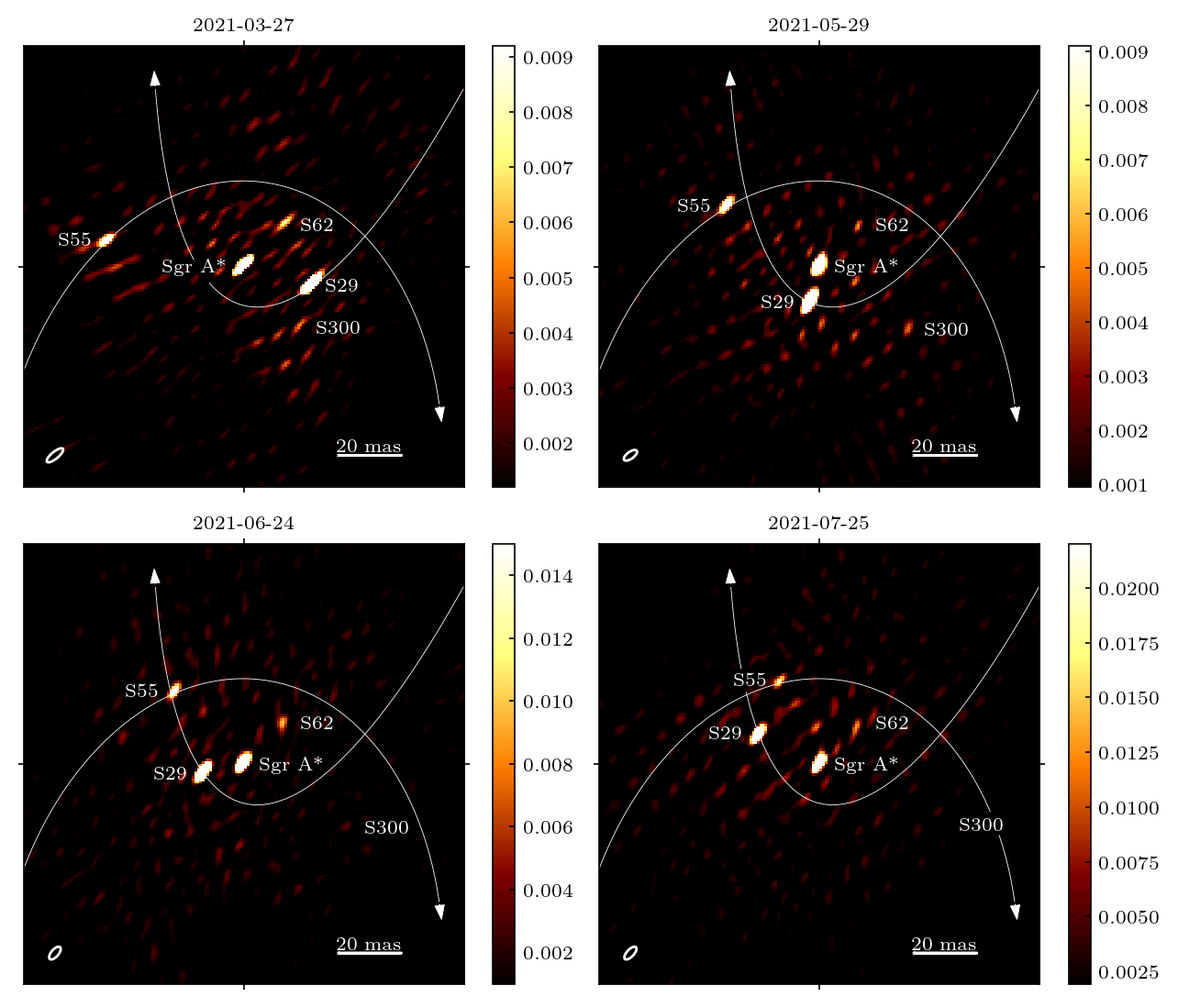}
        \caption{Representative image for each observing epoch obtained with CLEAN (north is up, east to the left, and the flux is normalized to S2). We show the model convolved with the CLEAN beam on top of the residual images and indicate the FWHM of the CLEAN beam in the bottom left corner. To obtain the model, we cleaned Sgr~A*, S29, S55, and S62; for the images from March to June, we also cleaned S300.}
        \label{fig:discussion-clean-images}
\end{figure*}

In addition to the heightened fiber damping, the large distance to the field center in combination with systematic effects can also affect the ability to detect S300 in July. In the context of our forward model, we can understand systematics as a mismatch between the predicted signal of the source and its actual effect on the data. It is expected that the quality of our modeling decreases for sources further away from the field center. The possible reasons for this include bandwidth smearing and optical aberrations in the instrument. Both effects become stronger further off axis, and, consequently, their modeling is more sensitive to approximations and calibration data, such as the bandpass shape or the aberration maps. It is important to note in this context, that Bayesian analyses in high dimensions are particularly vulnerable to model mismatches and that alternative tools can have a higher degree of robustness under such circumstances. We come back to this issue in Sect.~\ref{sec:discussion-clean}, when we compare the new imaging algorithm to CLEAN.

At this point, we have encountered several factors beyond $\left(u,v\right)$-coverage and data quality that can considerably impact the sensitivity of our images, such as the brightness of Sgr~A*, proximity to another faint source and the distance to the image center. Rather than giving a single estimate for the limiting magnitude, we therefore decided to highlight two bracketing values. Even under somewhat difficult circumstances -- in June, S300 is already seen to be outside the FWHM of the fiber and in the first mock test, the injected source is very close to S300 -- we are able to recover a source of at least $ m_\mathrm{K}\simeq 19.7$ magnitude when corrected for fiber damping. On the other hand, mock tests at position 2, and also the low noise levels reported in Appendix~\ref{appx:residual-images}, indicate that under favorable circumstances, we are able to push the sensitivity significantly beyond a magnitude of 20 with the newly developed \gr~imaging algorithm.

\section{Discussion}
\label{sec:discussion}

\subsection{Confirmation of S300 with CLEAN}
\label{sec:discussion-clean}

So far, the standard for deep imaging of the GC with GRAVITY has been set by CLEAN \citep{2021A&A...645A.127G}, and comparing both methods is important to judge the performance of our new \gr~algorithm. We therefore carried out an independent analysis of the 2021 data with CLEAN \citep[using the AIPS implementation, ][]{2003ASSL..285..109G}, which was already informed of the detection of S300 with \gr. In Fig.~\ref{fig:discussion-clean-images}, we provide a representative image for each observing epoch and in Table~\ref{tab:discussion-clean-data-summary}, we list the corresponding data sets.

\begin{figure*}
        \sidecaption
        \includegraphics[]{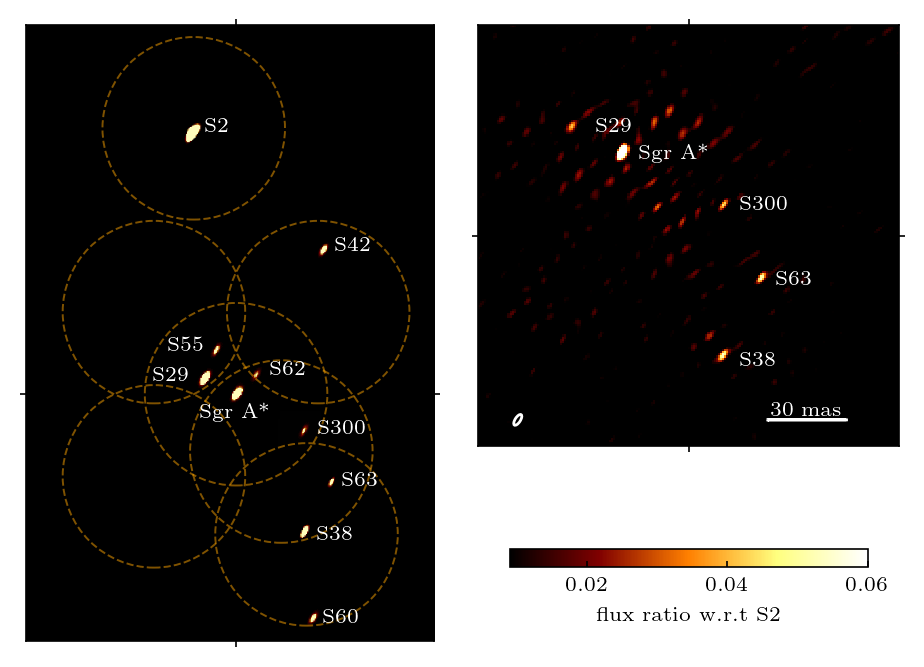}%
        \caption{Imaging results for the wide field data set obtained with CLEAN, where north is up, east to the left, and the flux is normalized to S2. In the left panel, we combine the CLEAN model from all pointings in the July 2021 mosaicing data set (see Table~{\ref{tab:data-observations-widefield}}) and indicate the pointing direction by a circle with $50\,\si{mas}$ radius. Apart from the expected stars, which are also summarized in Fig.~\ref{fig:res-offpointings-summary}, we do not detect any additional sources. In the right panel, we show the mid pointings from July 25, 26, and 27 (16 exposures in total) imaged with CLEAN. The image displays the model convolved with the CLEAN beam on top of the residuals and the beam size is indicated in the bottom left corner. To obtain the model, we have cleaned on Sgr~A*, S38, S63, S300, and S29.}
        \label{fig:discussion-clean-midpointing}
\end{figure*}

To successfully apply CLEAN to GRAVITY GC observations, a distinct procedure was developed and described in \cite{2021A&A...645A.127G}. Here, we adapt it to the changed field configuration in 2021. The major steps are laid out in the following.

Just as in 2019, we start by computing the coherent flux from the complex visibilities and the photometric flux observed at each telescope. To correct for flux variations (introduced e.g., by changes in air mass or the performance of the adaptive optics system), we interpolated the photometric flux of S2 across all frames collected for a night and normalized the coherent flux in the Sgr~A*-centered pointings by this value.

We then cleaned on Sgr~A* and S29 on an exposure-by-exposure basis. In particular, S29 is the brightest source in the field, but cleaning on Sgr~A* is important due to its flux variations, which would otherwise render the data mutually inconsistent. After this has been resolved, we combine all Sgr~A*- and S29-cleaned exposures from the night. In the resulting residual image, S55 is clearly visible and we jointly clean on the remains of Sgr~A* and S29, as well as on S55. At this point, S62 usually becomes apparent. For three of the four nights shown in Fig.~\ref{fig:discussion-clean-images}, it was the brightest residual, and it was only in the May imaging that it appeared as the second brightest. 

The images obtained with CLEAN (cf. Fig.~\ref{fig:discussion-clean-images}) agree very well with the \gr~results (cf. Fig.~\ref{fig:results-images}). Both methods recover an identical image structure that is dominated by the four established stars and matching source positions. Further, the CLEAN images confirm the detection of S300. For the March and May nights shown in Fig.~\ref{fig:discussion-clean-images}, after S62 has been cleaned, a bright residual becomes visible whose location is consistent with the \gr~discovery. It corresponds to the second-brightest residual on March 27 and the third-brightest on May 29. Also on the June night, a residual is visible at the expected S300 position; however, in this case, it is not among the few brightest ones. Here, fiber damping considerably impacts the ability to detect S300 in the CLEAN images. Bandwidth smearing, which becomes more severe for sources further away from the image center and is not modeled by CLEAN, can also diminish the sensitivity to S300 for the later 2021 observing runs. Finally, the images in Fig.~\ref{fig:discussion-clean-images} were obtained by cleaning on S300.

The most powerful confirmation of S300 in the CLEAN images, however, is provided by the July mid pointings in Fig.~\ref{fig:discussion-clean-midpointing}. The image is obtained from a total of 16 frames collected over three consecutive nights from July 25 to July 27, 2021. After cleaning on Sgr~A* in each individual exposure and combining all files, S38, S63, and S300 become apparent in the residual image, and we clean on the sources in this order. Subsequently, S29 can be recognized, which appears fainter in the mid pointings due to the larger fiber damping, and we also cleaned on it. Again, the CLEAN image shows excellent consistency with the \gr~result in Fig.~\ref{fig:results-offpointing-images}.

As in \cite{2021A&A...645A.127G}, we computed the root mean square (rms) value of the residual image in the central $74 \times 74\,\si{mas}$ for the Sgr~A* centered images, see Table~\ref{tab:discussion-clean-data-summary}. The rms of the complex visibilities and the residual image are directly related by Parseval's theorem, such that the latter quantity measures how well the model is able to reproduce the data. The flux in the CLEAN images as well as the rms values are normalized to S2, a $14.1$ magnitude star in K-band \citep{2017A&A...602A..94G}. Equivalent values for the \gr~results are estimated in Appendix~\ref{appx:residual-images} and are given in Table~\ref{appx:tab-noise-rms}.

\begin{table}
\caption{Summary of the data set from which the CLEAN images in Fig.~\ref{fig:discussion-clean-images} were obtained.}
\label{tab:discussion-clean-data-summary}
\centering
\begin{tabular}{c c c c}
Date  & $N_\mathrm{exp}$ & rms/S2 & \shortstack{corresponding $m_\mathrm{K}$}\\
\hline
2021-03-27 &  $12$ & $1.5\times10^{-3}$ & $21.2$ \\
2021-05-29 &  $20$ & $1.2\times10^{-3}$ & $21.4$ \\
2021-06-24 &  $32$ & $1.4\times10^{-3}$ & $21.2$ \\
2021-07-25 &  $28$ & $2.5\times10^{-3}$ & $20.6$
\end{tabular}
\tablefoot{Here, $N_\mathrm{exp}$ gives the number of exposures used for imaging. In addition, we also provide the rms in the residual image normalized to S2 and translate the value to a K-band magnitude in the final column.}
\end{table}

The low brightness of S300 in the June CLEAN residual image, where \gr~still manages a robust detection, already illustrates the increased sensitivity of the latter method. Furthermore, for the May and June observing nights, where we can directly compare the results in Table~\ref{tab:discussion-clean-data-summary} and Table~\ref{appx:tab-noise-rms}, \gr~improves the rms over the CLEAN result by $\Delta m_\mathrm{K} \simeq 0.3 - 0.4$. At this point, we want to emphasize that the comparison of residual images is rather unfavorable for \gr, which reconstructs the image from closure phases and visibility amplitudes. Residual images are only computed retroactively and require some additional alignment steps which are described in Appendix~\ref{appx:residual-images} and can increase the noise level by themselves.

Apart from being better able to describe instrumental effects, \gr~has another advantage. The residual images are essential within CLEAN to search for new sources. They do not, however, contain any information about the error bars in the data domain. Even if the residual visibilities were completely consistent with the Gaussian noise expectation, the inhomogeneous sampling pattern of the Fourier plane would introduce structures in the residual images. It can be challenging to judge whether a bright spot in the residual image is consistent with noise expectations or corresponds to a faint source. In contrast, \gr~judges the match between model and data directly in the domain of the data and thus can compare the residuals directly with the error bars of the individual data points.

On the other hand, a clear advantage of CLEAN is its speed and the significantly smaller computational demands in comparison to \gr. A full image reconstruction starting from data conversion with CLEAN only requires a single CPU and takes one to one and a half hours, which are dominated by data selection and inspecting the results rather than computation time. The \gr-code runs for about ten times as long on 16 CPUs, albeit once it has been initiated, it requires no human intervention.

Finally, we note that in some situations, CLEAN appears to be more robust to shortcomings of the response model. For the mid-pointing images with \gr~in Sect.~\ref{sec:results-mosicing}, we only used data from a single night. Combining multiple nights with \gr~works well for S2 and S38 pointings, but this significantly degrades the imaging solution of the mid pointings by introducing spurious sources, most likely due to the day-to-day movement of S29 and S55. While this further illustrates the heightened sensitivity of \gr, it also implies that additional work on the prior model will be required before one can combine exposures with fast moving sources from multiple nights. CLEAN, on the other hand, is able to image mid pointings from three consecutive nights jointly, without any artifacts from the movement of fainter stars affecting the structure of the brightest sources. In Fig.~{\ref{fig:discussion-clean-midpointing}}, we have Sgr~A*, S300, S63, and S38 all correctly recovered, while the fast-moving S29, which is subdominant to the total flux, appears slightly smeared out at an average position. On the other hand, S55, which is even fainter and further off axis than S29, cannot be recovered.

\subsection{Possible S300 positions in the Milky Way}
\label{sec:discussion-s300}
The observed S300 positions from March to July (cf. Table~\ref{tab:results-s300-pos-flux} and Sect.~\ref{sec:results-mosicing}) are fully consistent with a linear motion. Fitting for the angular velocity, we obtain:
\begin{align*}
v_\mathrm{RA}  = \left(-65.6 \pm 1.8\right)\,\si{mas/yr}\,,\\
v_\mathrm{Dec} = \left(-0.5 \pm 2.7\right)\,\si{mas/yr}\,.
\end{align*}
Its high angular velocity makes it very unlikely that S300 is a background star. In the following, we discuss possible options for describing its nature, namely, as a star located in the CG and a foreground star in the galactic disk.

If the star is at the same distance as the GC, its projected velocity with respect to the GC would be $v\simeq 2.6\times 10^3\,\si{\km/\s}$. The $2\sigma$ limit on the acceleration of S300 is $-23\,\mathrm{mas}/\mathrm{yr}^2$. This, in combination with requiring it to be gravitational bound to the black hole, limits the possible range of perpendicular coordinates to $\SI{50}{mas} \leq \left|z\right| \leq \SI{137}{mas}$. Furthermore, for each possible value of $z$ there is a maximum velocity $v_z$ beyond which the star would become unbound. Sampling uniformly from the allowed values for $z$ and their corresponding velocities, we can investigate the possible orbits of S300.

We find that the median orbit of this distribution has a  $171\,\mathrm{mas}$ semi-major axis, $0.6$ eccentricity, and a $26\,\mathrm{yr}$ orbital period. It thus perfectly fits into the distribution of S-stars. Assuming a similarly accurate position determination as for the 2021 imaging, a single good observing night from 2022 would allow to detect the acceleration of this median orbit. However, the distribution also contains more extreme solutions with larger semi-major axis and orbital period. For them, the apparently linear motion could continue throughout 2022. Even though, the continued observation of S300 over the coming year will allow to considerably constrain the family of allowed orbits.

Until then, in the absence of a significant acceleration detection, the possibility remains that S300 is a foreground star rather than being located in the GC. In this case, there would be little dust attenuation and S300 should also be detectable in the optical. We therefore checked the Gaia EDR3 catalog \citep{2021A&A...649A...1G}, which lists sources down to a G-band magnitude of $21$, but we do not obtain a match. In a circle with $4\,\si{arc~min}$ diameter around Sgr~A*, Gaia lists $285$ stars fainter than $m_\mathrm{G}>17$. If we use this number at face value to estimate the number density of stars towards the GC, the probability for a star crossing our $100 \times 100\,\si{mas}$ image is small $\simeq6\times 10^{-5}$.

Further, if S300 was a disk star, there would be only a narrow range of possible distances from the sun consistent with our observations. While the high angular velocity implies that S300 must be close, the linear motion and the absence of a parallax also impose a minimum distance. Taken together, in addition to the low probability of a foreground star crossing the narrow GRAVITY FOV, its distance to the sun is limited to $\mathcal{O}\left(\mathrm{kpc}\right)$. 

\section{Conclusions}
\label{sec:conclusions}
In this paper, we present GRAVITY-RESOLVE (\gr), a new imaging code, specifically tailored to GRAVITY observations of the Galactic Center (GC). The tool is based on a Bayesian interpretation of the imaging process and builds upon RESOLVE \citep{2021A&A...646A..84A, 2018arXiv180302174A}, an imaging tool for radio interferometry developed in the framework of information field theory \citep{2019AnP...53170017E}. In this context, we implemented an instrument model which accounts for all relevant effects in GRAVITY and developed a prior that is specifically designed for the GC and can, for instance, accommodate the variability of Sgr~A*. The posterior exploration was performed with Metric Gaussian Variational Inference \citep[MGVI, ][]{2019arXiv190111033K}.

We then applied \gr~to GC observations in 2021. The resulting images reveal a complicated structure, composed of several point sources of different brightness with a time-variable central object. We note that while our prior model is specifically tailored to the GC and only considers point sources, there are also methods available to model an extended emission within a similar framework \mbox{\citep[][]{2020arXiv200205218A, 2021A&A...646A..84A}}.

The stars S29 and S55 \citep{2017ApJ...837...30G} both pass their pericenters in 2021 and we detected them within a $50\,\mathrm{mas}$ radius from Sgr~A*. Their position in the images is also the starting point for astrometric model fitting which allows to determine the sources' positions to a very high accuracy of $\sim100\,\upmu\mathrm{as}$. We present a detailed study of the resulting GR orbits in a second publication \citep{mass_distribution}. Further, \gr~yields a very robust night-by-night detection of S62, a $m_\mathrm{K} = 18.9$ star that slowly approaches Sgr~A* and has already been found in GRAVITY GC images from 2019 obtained with CLEAN \citep{2021A&A...645A.127G}. None of the sources S29, S55, S62, and S300 identified in the GRAVITY observations matches a 9.9 year orbital period star as reported in \mbox{\cite{2020ApJ...889...61P, 2021ApJ...918...25P}}.

In addition to the known stars, a new source is apparent from the images which moves to the west at high angular speed $\simeq 66\,\si{mas/yr}$. It is detected at a high significance in the March, May, and June observations, but only dimly recognizable in the July Sgr~A*-centered exposures, where it is located at the largest distance from the image center. In July, however, we performed some dedicated mid pointings in which the GRAVITY fibers are offset from Sgr~A* and indeed recover the star at the expected location. 

The new source neither corresponds to any of the known S-stars \citep{2017ApJ...837...30G} nor is it present in the Gaia catalog \citep{2021A&A...649A...1G}, and we refer to it as S300. From the flux ratio with known stars in the field, particularly S29 and S38, we estimate that $m_\mathrm{K}\left(S300\right) = 19.0 - 19.3$. If fiber damping is taken into account, this source is at the detection limit estimated in \cite{2021A&A...645A.127G} in the March images and becomes dimmer during the rest of the year. The detection clearly demonstrates that \gr~can deliver significantly deeper images. With the knowledge of the S300 position, we also searched for it in 2021 CLEAN images and, indeed, we can identify a bright residual at the correct location in March, May, June, and the mid pointings.

The mid pointings are part of a larger mosaicing data set (cf. Table~\ref{tab:data-observations-widefield}), obtained to scan a wider field around Sgr A*. In the corresponding images (cf. Fig~\ref{fig:results-offpointing-images}), we can detect S38, S42, S60, and S63 in addition to the aforementioned sources. We use the opportunity to give an update on the orbital elements of all stars in Table~\ref{tab:results-orbital-elements}. These are based on astrometric fits to the data for which the imaging serves as a starting point.

Struggles to detect S300 in the July Sgr~A*-centered frames are partly due to fiber damping, but also the fact that our instrumental model becomes more sensitive to approximations; in addition, uncertainties in the calibration the farther away from the image center a source is located might play a role. To further assess the sensitivity of the new code, we performed a series of injection tests that we compared to the apparent magnitude of S300. As the images from May 22 demonstrate, a flare significantly reduces our sensitivity to faint sources. Under moderately difficult circumstances -- if the source is injected close to S300 or in June, where the off-axis separation already exceeds the fiber FWHM -- we are still able to robustly recover a source with an apparent magnitude of at least 19.7. Finally, the injection tests show that, under good circumstances, we are able to push the sensitivity significantly below a magnitude of 20 and we are even able to retrieve a  injected star of a magnitude of 21.0.

At present, a limiting factor to \gr~is that we have to scale the error bars by hand and to adjust that scaling by trial-and-error, making the application cumbersome. This can be improved in the future by implementing an automated error inference, such as in \cite{2019A&A...627A.134A}. With this method, we can also introduce more elaborate covariance matrices for the likelihood. Indeed, we find that the correlation structure of our residuals is similar to the one reported in \cite{2020A&A...644A.110K}. This analysis, which was carried out in the context of exo-planet observations with GRAVITY, demonstrated that an improved correlation model increases the achievable contrast. We see a similar potential in the imaging context, which would allow for an improvement in the convergence of the inference and further enhance the sensitivity beyond the capabilities that we have demonstrated with this publication.

\begin{acknowledgements}
We are very grateful to our funding agencies (MPG, ERC, CNRS [PNCG, PNGRAM], DFG, BMBF, Paris Observatory [CS, PhyFOG], Observatoire des Sciences de l’Univers de Grenoble, and the Fundação para a Ciência e Tecnologia), to ESO and the ESO/Paranal staff, and to the many scientific and technical staff members in our institutions, who helped to make GRAVITY a reality. This publication is based on observations collected at the European Southern Observatory under ESO programme 105.20B2.002. A.A.,  P.G. and V.C. were supported by Funda\c{c}\~{a}o para a Ci\^{e}ncia e a Tecnologia, with grants reference SFRH/BSAB/142940/2018, UIDB/00099/2020 and PTDC/FIS-AST/7002/2020. P.A.\ acknowledges the financial support by the German Federal Ministry of Education and Research (BMBF) under grant 05A17PB1 (Verbundprojekt D-MeerKAT). In the implementation of our code, we relied on the NIFTy and ducc0 Python packages and used the computing resources provided by MPCDF for our inferences.
\end{acknowledgements}

\bibliographystyle{aa}
\bibliography{gravity_imaging_paper}

\begin{thebibliography}{60}
\expandafter\ifx\csname natexlab\endcsname\relax\def\natexlab#1{#1}\fi

\bibitem[{{Arras} {et~al.}(2019{\natexlab{a}}){Arras}, {Baltac}, {Ensslin},
  {Frank}, {Hutschenreuter}, {Knollmueller}, {Leike}, {Newrzella}, {Platz},
  {Reinecke}, \& {Stadler}}]{2019ascl.soft03008A}
{Arras}, P., {Baltac}, M., {Ensslin}, T.~A., {et~al.} 2019{\natexlab{a}},
  {NIFTy5: Numerical Information Field Theory v5}

\bibitem[{{Arras} {et~al.}(2021{\natexlab{a}}){Arras}, {Bester}, {Perley},
  {Leike}, {Smirnov}, {Westermann}, \& {En{\ss}lin}}]{2021A&A...646A..84A}
{Arras}, P., {Bester}, H.~L., {Perley}, R.~A., {et~al.} 2021{\natexlab{a}},
  \aap, 646, A84

\bibitem[{{Arras} {et~al.}(2020){Arras}, {Frank}, {Haim}, {Knollm{\"u}ller},
  {Leike}, {Reinecke}, \& {En{\ss}lin}}]{2020arXiv200205218A}
{Arras}, P., {Frank}, P., {Haim}, P., {et~al.} 2020, arXiv e-prints,
  arXiv:2002.05218

\bibitem[{{Arras} {et~al.}(2019{\natexlab{b}}){Arras}, {Frank}, {Leike},
  {Westermann}, \& {En{\ss}lin}}]{2019A&A...627A.134A}
{Arras}, P., {Frank}, P., {Leike}, R., {Westermann}, R., \& {En{\ss}lin}, T.~A.
  2019{\natexlab{b}}, \aap, 627, A134

\bibitem[{Arras {et~al.}(2018)Arras, Knollrnüller, Junklewitz, \&
  Enßlin}]{2018arXiv180302174A}
Arras, P., Knollrnüller, J., Junklewitz, H., \& Enßlin, T.~A. 2018, in 2018
  26th European Signal Processing Conference (EUSIPCO), 2683--2687

\bibitem[{{Arras} {et~al.}(2021{\natexlab{b}}){Arras}, {Reinecke},
  {Westermann}, \& {En{\ss}in}}]{2021A&A...646A..58A}
{Arras}, P., {Reinecke}, M., {Westermann}, R., \& {En{\ss}in}, T.~A.
  2021{\natexlab{b}}, \aap, 646, A58

\bibitem[{{Baron} {et~al.}(2010){Baron}, {Monnier}, \&
  {Kloppenborg}}]{2010SPIE.7734E..2IB}
{Baron}, F., {Monnier}, J.~D., \& {Kloppenborg}, B. 2010, in Society of
  Photo-Optical Instrumentation Engineers (SPIE) Conference Series, Vol. 7734,
  Optical and Infrared Interferometry II, ed. W.~C. {Danchi}, F.~{Delplancke},
  \& J.~K. {Rajagopal}, 77342I

\bibitem[{{Bhatnagar} {et~al.}(2008){Bhatnagar}, {Cornwell}, {Golap}, \&
  {Uson}}]{2008A&A...487..419B}
{Bhatnagar}, S., {Cornwell}, T.~J., {Golap}, K., \& {Uson}, J.~M. 2008, \aap,
  487, 419

\bibitem[{{Blackburn} {et~al.}(2020){Blackburn}, {Pesce}, {Johnson}, {Wielgus},
  {Chael}, {Christian}, \& {Doeleman}}]{2020ApJ...894...31B}
{Blackburn}, L., {Pesce}, D.~W., {Johnson}, M.~D., {et~al.} 2020, \apj, 894, 31

\bibitem[{{Briggs} {et~al.}(1999){Briggs}, {Schwab}, \&
  {Sramek}}]{1999ASPC..180..127B}
{Briggs}, D.~S., {Schwab}, F.~R., \& {Sramek}, R.~A. 1999, in Astronomical
  Society of the Pacific Conference Series, Vol. 180, Synthesis Imaging in
  Radio Astronomy II, ed. G.~B. {Taylor}, C.~L. {Carilli}, \& R.~A. {Perley},
  127

\bibitem[{{Do} {et~al.}(2009){Do}, {Ghez}, {Morris}, {Yelda}, {Meyer}, {Lu},
  {Hornstein}, \& {Matthews}}]{2009ApJ...691.1021D}
{Do}, T., {Ghez}, A.~M., {Morris}, M.~R., {et~al.} 2009, \apj, 691, 1021

\bibitem[{{Do} {et~al.}(2019){Do}, {Hees}, {Ghez}, {Martinez}, {Chu}, {Jia},
  {Sakai}, {Lu}, {Gautam}, {O'Neil}, {Becklin}, {Morris}, {Matthews},
  {Nishiyama}, {Campbell}, {Chappell}, {Chen}, {Ciurlo}, {Dehghanfar},
  {Gallego-Cano}, {Kerzendorf}, {Lyke}, {Naoz}, {Saida}, {Sch{\"o}del},
  {Takahashi}, {Takamori}, {Witzel}, \& {Wizinowich}}]{2019Sci...365..664D}
{Do}, T., {Hees}, A., {Ghez}, A., {et~al.} 2019, Science, 365, 664

\bibitem[{{Do} {et~al.}(2013){Do}, {Lu}, {Ghez}, {Morris}, {Yelda}, {Martinez},
  {Wright}, \& {Matthews}}]{2013ApJ...764..154D}
{Do}, T., {Lu}, J.~R., {Ghez}, A.~M., {et~al.} 2013, \apj, 764, 154

\bibitem[{{Dodds-Eden} {et~al.}(2011){Dodds-Eden}, {Gillessen}, {Fritz},
  {Eisenhauer}, {Trippe}, {Genzel}, {Ott}, {Bartko}, {Pfuhl}, {Bower},
  {Goldwurm}, {Porquet}, {Trap}, \& {Yusef-Zadeh}}]{2011ApJ...728...37D}
{Dodds-Eden}, K., {Gillessen}, S., {Fritz}, T.~K., {et~al.} 2011, \apj, 728, 37

\bibitem[{{Dutt} \& {Rokhlin}(1993)}]{gridding-article}
{Dutt}, A. \& {Rokhlin}, V. 1993, SIAM Journal on Scientific Computing, 14,
  1368

\bibitem[{{Eckart} {et~al.}(2008){Eckart}, {Sch{\"o}del},
  {Garc{\'\i}a-Mar{\'\i}n}, {Witzel}, {Weiss}, {Baganoff}, {Morris}, {Bertram},
  {Dov{\v{c}}iak}, {Duschl}, {Karas}, {K{\"o}nig}, {Krichbaum}, {Krips},
  {Kunneriath}, {Lu}, {Markoff}, {Mauerhan}, {Meyer}, {Moultaka},
  {Mu{\v{z}}i{\'c}}, {Najarro}, {Pott}, {Schuster}, {Sjouwerman},
  {Straubmeier}, {Thum}, {Vogel}, {Wiesemeyer}, {Zamaninasab}, \&
  {Zensus}}]{2008A&A...492..337E}
{Eckart}, A., {Sch{\"o}del}, R., {Garc{\'\i}a-Mar{\'\i}n}, M., {et~al.} 2008,
  \aap, 492, 337

\bibitem[{{Eisenhauer} {et~al.}(2005){Eisenhauer}, {Genzel}, {Alexander},
  {Abuter}, {Paumard}, {Ott}, {Gilbert}, {Gillessen}, {Horrobin}, {Trippe},
  {Bonnet}, {Dumas}, {Hubin}, {Kaufer}, {Kissler-Patig}, {Monnet},
  {Str{\"o}bele}, {Szeifert}, {Eckart}, {Sch{\"o}del}, \&
  {Zucker}}]{2005ApJ...628..246E}
{Eisenhauer}, F., {Genzel}, R., {Alexander}, T., {et~al.} 2005, \apj, 628, 246

\bibitem[{{En{\ss}lin}(2019)}]{2019AnP...53170017E}
{En{\ss}lin}, T.~A. 2019, Annalen der Physik, 531, 1970017

\bibitem[{{Gaia Collaboration} {et~al.}(2021){Gaia Collaboration}, {Brown},
  {Vallenari}, {Prusti}, {de Bruijne}, {Babusiaux}, {Biermann}, {Creevey},
  {Evans}, {Eyer}, {Hutton}, {Jansen}, {Jordi}, {Klioner}, {Lammers},
  {Lindegren}, {Luri}, {Mignard}, {Panem}, {Pourbaix}, {Randich}, {Sartoretti},
  {Soubiran}, {Walton}, {Arenou}, {Bailer-Jones}, {Bastian}, {Cropper},
  {Drimmel}, {Katz}, {Lattanzi}, {van Leeuwen}, {Bakker}, {Cacciari},
  {Casta{\~n}eda}, {De Angeli}, {Ducourant}, {Fabricius}, {Fouesneau},
  {Fr{\'e}mat}, {Guerra}, {Guerrier}, {Guiraud}, {Jean-Antoine Piccolo},
  {Masana}, {Messineo}, {Mowlavi}, {Nicolas}, {Nienartowicz}, {Pailler},
  {Panuzzo}, {Riclet}, {Roux}, {Seabroke}, {Sordo}, {Tanga}, {Th{\'e}venin},
  {Gracia-Abril}, {Portell}, {Teyssier}, {Altmann}, {Andrae}, {Bellas-Velidis},
  {Benson}, {Berthier}, {Blomme}, {Brugaletta}, {Burgess}, {Busso}, {Carry},
  {Cellino}, {Cheek}, {Clementini}, {Damerdji}, {Davidson}, {Delchambre},
  {Dell'Oro}, {Fern{\'a}ndez-Hern{\'a}ndez}, {Galluccio}, {Garc{\'\i}a-Lario},
  {Garcia-Reinaldos}, {Gonz{\'a}lez-N{\'u}{\~n}ez}, {Gosset}, {Haigron},
  {Halbwachs}, {Hambly}, {Harrison}, {Hatzidimitriou}, {Heiter},
  {Hern{\'a}ndez}, {Hestroffer}, {Hodgkin}, {Holl}, {Jan{\ss}en}, {Jevardat de
  Fombelle}, {Jordan}, {Krone-Martins}, {Lanzafame}, {L{\"o}ffler}, {Lorca},
  {Manteiga}, {Marchal}, {Marrese}, {Moitinho}, {Mora}, {Muinonen}, {Osborne},
  {Pancino}, {Pauwels}, {Petit}, {Recio-Blanco}, {Richards}, {Riello},
  {Rimoldini}, {Robin}, {Roegiers}, {Rybizki}, {Sarro}, {Siopis}, {Smith},
  {Sozzetti}, {Ulla}, {Utrilla}, {van Leeuwen}, {van Reeven}, {Abbas}, {Abreu
  Aramburu}, {Accart}, {Aerts}, {Aguado}, {Ajaj}, {Altavilla}, {{\'A}lvarez},
  {{\'A}lvarez Cid-Fuentes}, {Alves}, {Anderson}, {Anglada Varela}, {Antoja},
  {Audard}, {Baines}, {Baker}, {Balaguer-N{\'u}{\~n}ez}, {Balbinot}, {Balog},
  {Barache}, {Barbato}, {Barros}, {Barstow}, {Bartolom{\'e}}, {Bassilana},
  {Bauchet}, {Baudesson-Stella}, {Becciani}, {Bellazzini}, {Bernet}, {Bertone},
  {Bianchi}, {Blanco-Cuaresma}, {Boch}, {Bombrun}, {Bossini}, {Bouquillon},
  {Bragaglia}, {Bramante}, {Breedt}, {Bressan}, {Brouillet}, {Bucciarelli},
  {Burlacu}, {Busonero}, {Butkevich}, {Buzzi}, {Caffau}, {Cancelliere},
  {C{\'a}novas}, {Cantat-Gaudin}, {Carballo}, {Carlucci}, {Carnerero},
  {Carrasco}, {Casamiquela}, {Castellani}, {Castro-Ginard}, {Castro Sampol},
  {Chaoul}, {Charlot}, {Chemin}, {Chiavassa}, {Cioni}, {Comoretto}, {Cooper},
  {Cornez}, {Cowell}, {Crifo}, {Crosta}, {Crowley}, {Dafonte}, {Dapergolas},
  {David}, {David}, {de Laverny}, {De Luise}, {De March}, {De Ridder}, {de
  Souza}, {de Teodoro}, {de Torres}, {del Peloso}, {del Pozo}, {Delbo},
  {Delgado}, {Delgado}, {Delisle}, {Di Matteo}, {Diakite}, {Diener},
  {Distefano}, {Dolding}, {Eappachen}, {Edvardsson}, {Enke}, {Esquej}, {Fabre},
  {Fabrizio}, {Faigler}, {Fedorets}, {Fernique}, {Fienga}, {Figueras},
  {Fouron}, {Fragkoudi}, {Fraile}, {Franke}, {Gai}, {Garabato},
  {Garcia-Gutierrez}, {Garc{\'\i}a-Torres}, {Garofalo}, {Gavras}, {Gerlach},
  {Geyer}, {Giacobbe}, {Gilmore}, {Girona}, {Giuffrida}, {Gomel}, {Gomez},
  {Gonzalez-Santamaria}, {Gonz{\'a}lez-Vidal}, {Granvik},
  {Guti{\'e}rrez-S{\'a}nchez}, {Guy}, {Hauser}, {Haywood}, {Helmi}, {Hidalgo},
  {Hilger}, {H{\l}adczuk}, {Hobbs}, {Holland}, {Huckle}, {Jasniewicz},
  {Jonker}, {Juaristi Campillo}, {Julbe}, {Karbevska}, {Kervella}, {Khanna},
  {Kochoska}, {Kontizas}, {Kordopatis}, {Korn}, {Kostrzewa-Rutkowska},
  {Kruszy{\'n}ska}, {Lambert}, {Lanza}, {Lasne}, {Le Campion}, {Le Fustec},
  {Lebreton}, {Lebzelter}, {Leccia}, {Leclerc}, {Lecoeur-Taibi}, {Liao},
  {Licata}, {Lindstr{\o}m}, {Lister}, {Livanou}, {Lobel}, {Madrero Pardo},
  {Managau}, {Mann}, {Marchant}, {Marconi}, {Marcos Santos}, {Marinoni},
  {Marocco}, {Marshall}, {Martin Polo}, {Mart{\'\i}n-Fleitas}, {Masip},
  {Massari}, {Mastrobuono-Battisti}, {Mazeh}, {McMillan}, {Messina},
  {Michalik}, {Millar}, {Mints}, {Molina}, {Molinaro}, {Moln{\'a}r},
  {Montegriffo}, {Mor}, {Morbidelli}, {Morel}, {Morris}, {Mulone}, {Munoz},
  {Muraveva}, {Murphy}, {Musella}, {Noval}, {Ord{\'e}novic}, {Orr{\`u}},
  {Osinde}, {Pagani}, {Pagano}, {Palaversa}, {Palicio}, {Panahi}, {Pawlak},
  {Pe{\~n}alosa Esteller}, {Penttil{\"a}}, {Piersimoni}, {Pineau}, {Plachy},
  {Plum}, {Poggio}, {Poretti}, {Poujoulet}, {Pr{\v{s}}a}, {Pulone}, {Racero},
  {Ragaini}, {Rainer}, {Raiteri}, {Rambaux}, {Ramos}, {Ramos-Lerate}, {Re
  Fiorentin}, {Regibo}, {Reyl{\'e}}, {Ripepi}, {Riva}, {Rixon}, {Robichon},
  {Robin}, {Roelens}, {Rohrbasser}, {Romero-G{\'o}mez}, {Rowell}, {Royer},
  {Rybicki}, {Sadowski}, {Sagrist{\`a} Sell{\'e}s}, {Sahlmann}, {Salgado},
  {Salguero}, {Samaras}, {Sanchez Gimenez}, {Sanna}, {Santove{\~n}a},
  {Sarasso}, {Schultheis}, {Sciacca}, {Segol}, {Segovia}, {S{\'e}gransan},
  {Semeux}, {Shahaf}, {Siddiqui}, {Siebert}, {Siltala}, {Slezak}, {Smart},
  {Solano}, {Solitro}, {Souami}, {Souchay}, {Spagna}, {Spoto}, {Steele},
  {Steidelm{\"u}ller}, {Stephenson}, {S{\"u}veges}, {Szabados}, {Szegedi-Elek},
  {Taris}, {Tauran}, {Taylor}, {Teixeira}, {Thuillot}, {Tonello}, {Torra},
  {Torra}, {Turon}, {Unger}, {Vaillant}, {van Dillen}, {Vanel}, {Vecchiato},
  {Viala}, {Vicente}, {Voutsinas}, {Weiler}, {Wevers}, {Wyrzykowski}, {Yoldas},
  {Yvard}, {Zhao}, {Zorec}, {Zucker}, {Zurbach}, \&
  {Zwitter}}]{2021A&A...649A...1G}
{Gaia Collaboration}, {Brown}, A.~G.~A., {Vallenari}, A., {et~al.} 2021, \aap,
  649, A1

\bibitem[{{Gallego-Cano} {et~al.}(2018){Gallego-Cano}, {Sch{\"o}del}, {Dong},
  {Nogueras-Lara}, {Gallego-Calvente}, {Amaro-Seoane}, \&
  {Baumgardt}}]{2018A&A...609A..26G}
{Gallego-Cano}, E., {Sch{\"o}del}, R., {Dong}, H., {et~al.} 2018, \aap, 609,
  A26

\bibitem[{{Genzel} {et~al.}(2010){Genzel}, {Eisenhauer}, \&
  {Gillessen}}]{2010RvMP...82.3121G}
{Genzel}, R., {Eisenhauer}, F., \& {Gillessen}, S. 2010, Reviews of Modern
  Physics, 82, 3121

\bibitem[{{Genzel} {et~al.}(2003{\natexlab{a}}){Genzel}, {Sch{\"o}del}, {Ott},
  {Eckart}, {Alexander}, {Lacombe}, {Rouan}, \&
  {Aschenbach}}]{2003Natur.425..934G}
{Genzel}, R., {Sch{\"o}del}, R., {Ott}, T., {et~al.} 2003{\natexlab{a}}, \nat,
  425, 934

\bibitem[{{Genzel} {et~al.}(2003{\natexlab{b}}){Genzel}, {Sch{\"o}del}, {Ott},
  {Eisenhauer}, {Hofmann}, {Lehnert}, {Eckart}, {Alexander}, {Sternberg},
  {Lenzen}, {Cl{\'e}net}, {Lacombe}, {Rouan}, {Renzini}, \&
  {Tacconi-Garman}}]{2003ApJ...594..812G}
{Genzel}, R., {Sch{\"o}del}, R., {Ott}, T., {et~al.} 2003{\natexlab{b}}, \apj,
  594, 812

\bibitem[{{Ghez} {et~al.}(2004){Ghez}, {Wright}, {Matthews}, {Thompson}, {Le
  Mignant}, {Tanner}, {Hornstein}, {Morris}, {Becklin}, \&
  {Soifer}}]{2004ApJ...601L.159G}
{Ghez}, A.~M., {Wright}, S.~A., {Matthews}, K., {et~al.} 2004, \apjl, 601, L159

\bibitem[{{Gillessen} {et~al.}(2006){Gillessen}, {Eisenhauer}, {Quataert},
  {Genzel}, {Paumard}, {Trippe}, {Ott}, {Abuter}, {Eckart}, {Lagage},
  {Lehnert}, {Tacconi}, \& {Martins}}]{2006ApJ...640L.163G}
{Gillessen}, S., {Eisenhauer}, F., {Quataert}, E., {et~al.} 2006, \apjl, 640,
  L163

\bibitem[{{Gillessen} {et~al.}(2009){Gillessen}, {Eisenhauer}, {Trippe},
  {Alexander}, {Genzel}, {Martins}, \& {Ott}}]{2009ApJ...692.1075G}
{Gillessen}, S., {Eisenhauer}, F., {Trippe}, S., {et~al.} 2009, \apj, 692, 1075

\bibitem[{{Gillessen} {et~al.}(2017){Gillessen}, {Plewa}, {Eisenhauer}, {Sari},
  {Waisberg}, {Habibi}, {Pfuhl}, {George}, {Dexter}, {von Fellenberg}, {Ott},
  \& {Genzel}}]{2017ApJ...837...30G}
{Gillessen}, S., {Plewa}, P.~M., {Eisenhauer}, F., {et~al.} 2017, \apj, 837, 30

\bibitem[{{Gravity Collaboration}(2021)}]{mass_distribution}
{Gravity Collaboration}. 2021, \aap

\bibitem[{{Gravity Collaboration} {et~al.}(2017){Gravity Collaboration},
  {Abuter}, {Accardo}, {Amorim}, {Anugu}, {{\'A}vila}, {Azouaoui}, {Benisty},
  {Berger}, {Blind}, {Bonnet}, {Bourget}, {Brandner}, {Brast}, {Buron},
  {Burtscher}, {Cassaing}, {Chapron}, {Choquet}, {Cl{\'e}net}, {Collin},
  {Coud{\'e} Du Foresto}, {de Wit}, {de Zeeuw}, {Deen},
  {Delplancke-Str{\"o}bele}, {Dembet}, {Derie}, {Dexter}, {Duvert}, {Ebert},
  {Eckart}, {Eisenhauer}, {Esselborn}, {F{\'e}dou}, {Finger}, {Garcia}, {Garcia
  Dabo}, {Garcia Lopez}, {Gendron}, {Genzel}, {Gillessen}, {Gonte}, {Gordo},
  {Grould}, {Gr{\"o}zinger}, {Guieu}, {Haguenauer}, {Hans}, {Haubois}, {Haug},
  {Haussmann}, {Henning}, {Hippler}, {Horrobin}, {Huber}, {Hubert}, {Hubin},
  {Hummel}, {Jakob}, {Janssen}, {Jochum}, {Jocou}, {Kaufer}, {Kellner},
  {Kendrew}, {Kern}, {Kervella}, {Kiekebusch}, {Klein}, {Kok}, {Kolb}, {Kulas},
  {Lacour}, {Lapeyr{\`e}re}, {Lazareff}, {Le Bouquin}, {L{\`e}na}, {Lenzen},
  {L{\'e}v{\^e}que}, {Lippa}, {Magnard}, {Mehrgan}, {Mellein}, {M{\'e}rand},
  {Moreno-Ventas}, {Moulin}, {M{\"u}ller}, {M{\"u}ller}, {Neumann}, {Oberti},
  {Ott}, {Pallanca}, {Panduro}, {Pasquini}, {Paumard}, {Percheron}, {Perraut},
  {Perrin}, {Pfl{\"u}ger}, {Pfuhl}, {Phan Duc}, {Plewa}, {Popovic}, {Rabien},
  {Ram{\'\i}rez}, {Ramos}, {Rau}, {Riquelme}, {Rohloff}, {Rousset},
  {Sanchez-Bermudez}, {Scheithauer}, {Sch{\"o}ller}, {Schuhler}, {Spyromilio},
  {Straubmeier}, {Sturm}, {Suarez}, {Tristram}, {Ventura}, {Vincent},
  {Waisberg}, {Wank}, {Weber}, {Wieprecht}, {Wiest}, {Wiezorrek}, {Wittkowski},
  {Woillez}, {Wolff}, {Yazici}, {Ziegler}, \& {Zins}}]{2017A&A...602A..94G}
{Gravity Collaboration}, {Abuter}, R., {Accardo}, M., {et~al.} 2017, \aap, 602,
  A94

\bibitem[{{Gravity Collaboration} {et~al.}(2018{\natexlab{a}}){Gravity
  Collaboration}, {Abuter}, {Amorim}, {Anugu}, {Baub{\"o}ck}, {Benisty},
  {Berger}, {Blind}, {Bonnet}, {Brandner}, {Buron}, {Collin}, {Chapron},
  {Cl{\'e}net}, {Coud{\'e} Du Foresto}, {de Zeeuw}, {Deen},
  {Delplancke-Str{\"o}bele}, {Dembet}, {Dexter}, {Duvert}, {Eckart},
  {Eisenhauer}, {Finger}, {F{\"o}rster Schreiber}, {F{\'e}dou}, {Garcia},
  {Garcia Lopez}, {Gao}, {Gendron}, {Genzel}, {Gillessen}, {Gordo}, {Habibi},
  {Haubois}, {Haug}, {Hau{\ss}mann}, {Henning}, {Hippler}, {Horrobin},
  {Hubert}, {Hubin}, {Jimenez Rosales}, {Jochum}, {Jocou}, {Kaufer}, {Kellner},
  {Kendrew}, {Kervella}, {Kok}, {Kulas}, {Lacour}, {Lapeyr{\`e}re}, {Lazareff},
  {Le Bouquin}, {L{\'e}na}, {Lippa}, {Lenzen}, {M{\'e}rand}, {M{\"u}ler},
  {Neumann}, {Ott}, {Palanca}, {Paumard}, {Pasquini}, {Perraut}, {Perrin},
  {Pfuhl}, {Plewa}, {Rabien}, {Ram{\'\i}rez}, {Ramos}, {Rau},
  {Rodr{\'\i}guez-Coira}, {Rohloff}, {Rousset}, {Sanchez-Bermudez},
  {Scheithauer}, {Sch{\"o}ller}, {Schuler}, {Spyromilio}, {Straub},
  {Straubmeier}, {Sturm}, {Tacconi}, {Tristram}, {Vincent}, {von Fellenberg},
  {Wank}, {Waisberg}, {Widmann}, {Wieprecht}, {Wiest}, {Wiezorrek}, {Woillez},
  {Yazici}, {Ziegler}, \& {Zins}}]{2018A&A...615L..15G}
{Gravity Collaboration}, {Abuter}, R., {Amorim}, A., {et~al.}
  2018{\natexlab{a}}, \aap, 615, L15

\bibitem[{{Gravity Collaboration} {et~al.}(2020{\natexlab{a}}){Gravity
  Collaboration}, {Abuter}, {Amorim}, {Baub{\"o}ck}, {Berger}, {Bonnet},
  {Brandner}, {Cardoso}, {Cl{\'e}net}, {de Zeeuw}, {Dallilar}, {Dexter},
  {Eckart}, {Eisenhauer}, {F{\"o}rster Schreiber}, {Garcia}, {Gao}, {Gendron},
  {Genzel}, {Gillessen}, {Habibi}, {Haubois}, {Henning}, {Hippler}, {Horrobin},
  {Jim{\'e}nez-Rosales}, {Jochum}, {Jocou}, {Kaufer}, {Kervella}, {Lacour},
  {Lapeyr{\`e}re}, {Le Bouquin}, {L{\'e}na}, {Nowak}, {Ott}, {Paumard},
  {Perraut}, {Perrin}, {Pfuhl}, {Ponti}, {Rodriguez Coira}, {Shangguan},
  {Scheithauer}, {Stadler}, {Straub}, {Straubmeier}, {Sturm}, {Tacconi},
  {Vincent}, {von Fellenberg}, {Waisberg}, {Widmann}, {Wieprecht}, {Wiezorrek},
  {Woillez}, {Yazici}, \& {Zins}}]{2020A&A...638A...2G}
{Gravity Collaboration}, {Abuter}, R., {Amorim}, A., {et~al.}
  2020{\natexlab{a}}, \aap, 638, A2

\bibitem[{{Gravity Collaboration} {et~al.}(2020{\natexlab{b}}){Gravity
  Collaboration}, {Abuter}, {Amorim}, {Baub{\"o}ck}, {Berger}, {Bonnet},
  {Brandner}, {Cardoso}, {Cl{\'e}net}, {de Zeeuw}, {Dexter}, {Eckart},
  {Eisenhauer}, {F{\"o}rster Schreiber}, {Garcia}, {Gao}, {Gendron}, {Genzel},
  {Gillessen}, {Habibi}, {Haubois}, {Henning}, {Hippler}, {Horrobin},
  {Jim{\'e}nez-Rosales}, {Jochum}, {Jocou}, {Kaufer}, {Kervella}, {Lacour},
  {Lapeyr{\`e}re}, {Le Bouquin}, {L{\'e}na}, {Nowak}, {Ott}, {Paumard},
  {Perraut}, {Perrin}, {Pfuhl}, {Rodr{\'\i}guez-Coira}, {Shangguan},
  {Scheithauer}, {Stadler}, {Straub}, {Straubmeier}, {Sturm}, {Tacconi},
  {Vincent}, {von Fellenberg}, {Waisberg}, {Widmann}, {Wieprecht}, {Wiezorrek},
  {Woillez}, {Yazici}, \& {Zins}}]{2020A&A...636L...5G}
{Gravity Collaboration}, {Abuter}, R., {Amorim}, A., {et~al.}
  2020{\natexlab{b}}, \aap, 636, L5

\bibitem[{{Gravity Collaboration} {et~al.}(2018{\natexlab{b}}){Gravity
  Collaboration}, {Abuter}, {Amorim}, {Baub{\"o}ck}, {Berger}, {Bonnet},
  {Brandner}, {Cl{\'e}net}, {Coud{\'e} Du Foresto}, {de Zeeuw}, {Deen},
  {Dexter}, {Duvert}, {Eckart}, {Eisenhauer}, {F{\"o}rster Schreiber},
  {Garcia}, {Gao}, {Gendron}, {Genzel}, {Gillessen}, {Guajardo}, {Habibi},
  {Haubois}, {Henning}, {Hippler}, {Horrobin}, {Huber}, {Jim{\'e}nez-Rosales},
  {Jocou}, {Kervella}, {Lacour}, {Lapeyr{\`e}re}, {Lazareff}, {Le Bouquin},
  {L{\'e}na}, {Lippa}, {Ott}, {Panduro}, {Paumard}, {Perraut}, {Perrin},
  {Pfuhl}, {Plewa}, {Rabien}, {Rodr{\'\i}guez-Coira}, {Rousset}, {Sternberg},
  {Straub}, {Straubmeier}, {Sturm}, {Tacconi}, {Vincent}, {von Fellenberg},
  {Waisberg}, {Widmann}, {Wieprecht}, {Wiezorrek}, {Woillez}, \&
  {Yazici}}]{2018A&A...618L..10G}
{Gravity Collaboration}, {Abuter}, R., {Amorim}, A., {et~al.}
  2018{\natexlab{b}}, \aap, 618, L10

\bibitem[{{Gravity Collaboration} {et~al.}(2021{\natexlab{a}}){Gravity
  Collaboration}, {Abuter}, {Amorim}, {Baub{\"o}ck}, {Berger}, {Bonnet},
  {Brandner}, {Cl{\'e}net}, {Dallilar}, {Davies}, {de Zeeuw}, {Dexter},
  {Drescher}, {Eisenhauer}, {F{\"o}rster Schreiber}, {Garcia}, {Gao},
  {Gendron}, {Genzel}, {Gillessen}, {Habibi}, {Haubois}, {Hei{\ss}el},
  {Henning}, {Hippler}, {Horrobin}, {Jim{\'e}nez-Rosales}, {Jochum}, {Jocou},
  {Kaufer}, {Kervella}, {Lacour}, {Lapeyr{\`e}re}, {Le Bouquin}, {L{\'e}na},
  {Lutz}, {Nowak}, {Ott}, {Paumard}, {Perraut}, {Perrin}, {Pfuhl}, {Rabien},
  {Rodr{\'\i}guez-Coira}, {Shangguan}, {Shimizu}, {Scheithauer}, {Stadler},
  {Straub}, {Straubmeier}, {Sturm}, {Tacconi}, {Vincent}, {von Fellenberg},
  {Waisberg}, {Widmann}, {Wieprecht}, {Wiezorrek}, {Woillez}, {Yazici}, \&
  {Zins}}]{2021A&A...645A.127G}
{Gravity Collaboration}, {Abuter}, R., {Amorim}, A., {et~al.}
  2021{\natexlab{a}}, \aap, 645, A127

\bibitem[{{Gravity Collaboration} {et~al.}(2021{\natexlab{b}}){Gravity
  Collaboration}, {Abuter}, {Amorim}, {Baub{\"o}ck}, {Berger}, {Bonnet},
  {Brandner}, {Cl{\'e}net}, {Davies}, {de Zeeuw}, {Dexter}, {Dallilar},
  {Drescher}, {Eckart}, {Eisenhauer}, {F{\"o}rster Schreiber}, {Garcia}, {Gao},
  {Gendron}, {Genzel}, {Gillessen}, {Habibi}, {Haubois}, {Hei{\ss}el},
  {Henning}, {Hippler}, {Horrobin}, {Jim{\'e}nez-Rosales}, {Jochum}, {Jocou},
  {Kaufer}, {Kervella}, {Lacour}, {Lapeyr{\`e}re}, {Le Bouquin}, {L{\'e}na},
  {Lutz}, {Nowak}, {Ott}, {Paumard}, {Perraut}, {Perrin}, {Pfuhl}, {Rabien},
  {Rodr{\'\i}guez-Coira}, {Shangguan}, {Shimizu}, {Scheithauer}, {Stadler},
  {Straub}, {Straubmeier}, {Sturm}, {Tacconi}, {Vincent}, {von Fellenberg},
  {Waisberg}, {Widmann}, {Wieprecht}, {Wiezorrek}, {Woillez}, {Yazici},
  {Young}, \& {Zins}}]{2021A&A...647A..59G}
{Gravity Collaboration}, {Abuter}, R., {Amorim}, A., {et~al.}
  2021{\natexlab{b}}, \aap, 647, A59

\bibitem[{{Gravity Collaboration} {et~al.}(2020{\natexlab{c}}){Gravity
  Collaboration}, {Baub{\"o}ck}, {Dexter}, {Abuter}, {Amorim}, {Berger},
  {Bonnet}, {Brandner}, {Cl{\'e}net}, {Coud{\'e} Du Foresto}, {de Zeeuw},
  {Duvert}, {Eckart}, {Eisenhauer}, {F{\"o}rster Schreiber}, {Gao}, {Garcia},
  {Gendron}, {Genzel}, {Gerhard}, {Gillessen}, {Habibi}, {Haubois}, {Henning},
  {Hippler}, {Horrobin}, {Jim{\'e}nez-Rosales}, {Jocou}, {Kervella}, {Lacour},
  {Lapeyr{\`e}re}, {Le Bouquin}, {L{\'e}na}, {Ott}, {Paumard}, {Perraut},
  {Perrin}, {Pfuhl}, {Rabien}, {Rodriguez Coira}, {Rousset}, {Scheithauer},
  {Stadler}, {Sternberg}, {Straub}, {Straubmeier}, {Sturm}, {Tacconi},
  {Vincent}, {von Fellenberg}, {Waisberg}, {Widmann}, {Wieprecht}, {Wiezorrek},
  {Woillez}, \& {Yazici}}]{2020A&A...635A.143G}
{Gravity Collaboration}, {Baub{\"o}ck}, M., {Dexter}, J., {et~al.}
  2020{\natexlab{c}}, \aap, 635, A143

\bibitem[{{Gravity Collaboration} {et~al.}(2020{\natexlab{d}}){Gravity
  Collaboration}, {Jim{\'e}nez-Rosales}, {Dexter}, {Widmann}, {Baub{\"o}ck},
  {Abuter}, {Amorim}, {Berger}, {Bonnet}, {Brandner}, {Cl{\'e}net}, {de Zeeuw},
  {Eckart}, {Eisenhauer}, {F{\"o}rster Schreiber}, {Garcia}, {Gao}, {Gendron},
  {Genzel}, {Gillessen}, {Habibi}, {Haubois}, {Hei{\ss}el}, {Henning},
  {Hippler}, {Horrobin}, {Jochum}, {Jocou}, {Kaufer}, {Kervella}, {Lacour},
  {Lapeyr{\`e}re}, {Le Bouquin}, {L{\'e}na}, {Nowak}, {Ott}, {Paumard},
  {Perraut}, {Perrin}, {Pfuhl}, {Rodr{\'\i}guez-Coira}, {Shangguan},
  {Scheithauer}, {Stadler}, {Straub}, {Straubmeier}, {Sturm}, {Tacconi},
  {Vincent}, {von Fellenberg}, {Waisberg}, {Wieprecht}, {Wiezorrek}, {Woillez},
  {Yazici}, \& {Zins}}]{2020A&A...643A..56G}
{Gravity Collaboration}, {Jim{\'e}nez-Rosales}, A., {Dexter}, J., {et~al.}
  2020{\natexlab{d}}, \aap, 643, A56

\bibitem[{{Greisen}(2003)}]{2003ASSL..285..109G}
{Greisen}, E.~W. 2003, {AIPS, the VLA, and the VLBA}, ed. A.~{Heck}, Vol. 285,
  109

\bibitem[{{Guyon}(2002)}]{2002A&A...387..366G}
{Guyon}, O. 2002, \aap, 387, 366

\bibitem[{{H{\"o}gbom}(1974)}]{1974A&AS...15..417H}
{H{\"o}gbom}, J.~A. 1974, \aaps, 15, 417

\bibitem[{{Kammerer} {et~al.}(2020){Kammerer}, {M{\'e}rand}, {Ireland}, \&
  {Lacour}}]{2020A&A...644A.110K}
{Kammerer}, J., {M{\'e}rand}, A., {Ireland}, M.~J., \& {Lacour}, S. 2020, \aap,
  644, A110

\bibitem[{{Knollm{\"u}ller} \& {En{\ss}lin}(2019)}]{2019arXiv190111033K}
{Knollm{\"u}ller}, J. \& {En{\ss}lin}, T.~A. 2019, arXiv e-prints,
  arXiv:1901.11033

\bibitem[{{Merritt} {et~al.}(2010){Merritt}, {Alexander}, {Mikkola}, \&
  {Will}}]{2010PhRvD..81f2002M}
{Merritt}, D., {Alexander}, T., {Mikkola}, S., \& {Will}, C.~M. 2010, \prd, 81,
  062002

\bibitem[{{Meyer} {et~al.}(2012){Meyer}, {Ghez}, {Sch{\"o}del}, {Yelda},
  {Boehle}, {Lu}, {Do}, {Morris}, {Becklin}, \&
  {Matthews}}]{2012Sci...338...84M}
{Meyer}, L., {Ghez}, A.~M., {Sch{\"o}del}, R., {et~al.} 2012, Science, 338, 84

\bibitem[{{Pei{\ss}ker} {et~al.}(2021){Pei{\ss}ker}, {Eckart}, \&
  {Ali}}]{2021ApJ...918...25P}
{Pei{\ss}ker}, F., {Eckart}, A., \& {Ali}, B. 2021, \apj, 918, 25

\bibitem[{{Pei{\ss}ker} {et~al.}(2020){Pei{\ss}ker}, {Eckart}, \&
  {Parsa}}]{2020ApJ...889...61P}
{Pei{\ss}ker}, F., {Eckart}, A., \& {Parsa}, M. 2020, \apj, 889, 61

\bibitem[{{Perrin} \& {Woillez}(2019)}]{2019A&A...625A..48P}
{Perrin}, G. \& {Woillez}, J. 2019, \aap, 625, A48

\bibitem[{{Rogers} {et~al.}(1974){Rogers}, {Hinteregger}, {Whitney},
  {Counselman}, {Shapiro}, {Wittels}, {Klemperer}, {Warnock}, {Clark},
  {Hutton}, {Marandino}, {Ronnang}, {Rydbeck}, \&
  {Niell}}]{1974ApJ...193..293R}
{Rogers}, A.~E.~E., {Hinteregger}, H.~F., {Whitney}, A.~R., {et~al.} 1974,
  \apj, 193, 293

\bibitem[{{Shahzamanian} {et~al.}(2015){Shahzamanian}, {Eckart}, {Valencia-S.},
  {Witzel}, {Zamaninasab}, {Sabha}, {Garc{\'\i}a-Mar{\'\i}n}, {Karas},
  {Karssen}, {Borkar}, {Dov{\v{c}}iak}, {Kunneriath}, {Bursa}, {Buchholz},
  {Moultaka}, \& {Straubmeier}}]{2015A&A...576A..20S}
{Shahzamanian}, B., {Eckart}, A., {Valencia-S.}, M., {et~al.} 2015, \aap, 576,
  A20

\bibitem[{{Shaklan} \& {Roddier}(1988)}]{1988ApOpt..27.2334S}
{Shaklan}, S. \& {Roddier}, F. 1988, \ao, 27, 2334

\bibitem[{{Smirnov}(2011)}]{2011A&A...527A.106S}
{Smirnov}, O.~M. 2011, \aap, 527, A106

\bibitem[{{Thi{\'e}baut}(2008)}]{2008SPIE.7013E..1IT}
{Thi{\'e}baut}, E. 2008, in Society of Photo-Optical Instrumentation Engineers
  (SPIE) Conference Series, Vol. 7013, Optical and Infrared Interferometry, ed.
  M.~{Sch{\"o}ller}, W.~C. {Danchi}, \& F.~{Delplancke}, 70131I

\bibitem[{{Trippe} {et~al.}(2007){Trippe}, {Paumard}, {Ott}, {Gillessen},
  {Eisenhauer}, {Martins}, \& {Genzel}}]{2007MNRAS.375..764T}
{Trippe}, S., {Paumard}, T., {Ott}, T., {et~al.} 2007, \mnras, 375, 764

\bibitem[{{van Cittert}(1934)}]{1934Phy.....1..201V}
{van Cittert}, P.~H. 1934, Physica, 1, 201

\bibitem[{{Waisberg} {et~al.}(2018){Waisberg}, {Dexter}, {Gillessen}, {Pfuhl},
  {Eisenhauer}, {Plewa}, {Baub{\"o}ck}, {Jimenez-Rosales}, {Habibi}, {Ott},
  {von Fellenberg}, {Gao}, {Widmann}, \& {Genzel}}]{2018MNRAS.476.3600W}
{Waisberg}, I., {Dexter}, J., {Gillessen}, S., {et~al.} 2018, \mnras, 476, 3600

\bibitem[{{Witzel} {et~al.}(2018){Witzel}, {Martinez}, {Hora}, {Willner},
  {Morris}, {Gammie}, {Becklin}, {Ashby}, {Baganoff}, {Carey}, {Do}, {Fazio},
  {Ghez}, {Glaccum}, {Haggard}, {Herrero-Illana}, {Ingalls}, {Narayan}, \&
  {Smith}}]{2018ApJ...863...15W}
{Witzel}, G., {Martinez}, G., {Hora}, J., {et~al.} 2018, \apj, 863, 15

\bibitem[{{Zamaninasab} {et~al.}(2010){Zamaninasab}, {Eckart}, {Witzel},
  {Dovciak}, {Karas}, {Sch{\"o}del}, {Gie{\ss}{\"u}bel}, {Bremer},
  {Garc{\'\i}a-Mar{\'\i}n}, {Kunneriath}, {Mu{\v{z}}i{\'c}}, {Nishiyama},
  {Sabha}, {Straubmeier}, \& {Zensus}}]{2010A&A...510A...3Z}
{Zamaninasab}, M., {Eckart}, A., {Witzel}, G., {et~al.} 2010, \aap, 510, A3

\bibitem[{{Zernike}(1938)}]{1938Phy.....5..785Z}
{Zernike}, F. 1938, Physica, 5, 785

\bibitem[{{Zhang} \& {Iorio}(2017)}]{2017ApJ...834..198Z}
{Zhang}, F. \& {Iorio}, L. 2017, \apj, 834, 198

\bibitem[{{Zucker} {et~al.}(2006){Zucker}, {Alexander}, {Gillessen},
  {Eisenhauer}, \& {Genzel}}]{2006ApJ...639L..21Z}
{Zucker}, S., {Alexander}, T., {Gillessen}, S., {Eisenhauer}, F., \& {Genzel},
  R. 2006, \apjl, 639, L21

\end{thebibliography}

\begin{appendix}
\section{Efficient response implementation}
\label{appx:response}
In the presence of static instrumental optical aberrations \citep{2021A&A...647A..59G} and bandwidth smearing, the simplified response function in Eq.~(\ref{eq: method-visibilities}) is generalized to:
\begin{align}
v_{ij}\left(\vec{u}\right) = \frac
{\int\dd\lambda\, P_{\lambda_0}\left(\lambda\right)\int\dd\vec{s}\, \Pi_i\left(\vec{s}\right)\Pi_j^*\left(\vec{s}\right)\, I\left(\vec{s},\lambda\right) \ee^{-2\uppi i \vec{b}\cdot\vec{s}/\lambda }}
{\prod_{x=i,j}\sqrt{\int \dd\lambda~ P_{\lambda_0}\left(\lambda\right) \int\dd\vec{s}~\left|\Pi_x\left(\vec{s}\right)\right|^2 I\left(\vec{s},\lambda\right) }}\,,
\label{eq:appx-response-full}
\end{align}
where $i$ and $j$ label the two telescopes forming the baseline, $\vec{b,}$ and the Fourier coordinate is evaluated at the central wavelength $\lambda_0$, that is, $\vec{u}=\vec{b}/\lambda_0$. The spectral bandpass is normalized such that $\int\dd\lambda~P_{\lambda_0}\left(\lambda\right) = 1$. Finally, $\Pi_{i/j}$ summarizes the phase maps, $\phi_{i/j}$, and amplitude maps, $A_{i/j}$, of each telescope and is given by
\begin{equation}
\Pi_i\left(\vec{s}\right) = A_i\left(\vec{s}+\vec{\delta}_i\right)\, \ee^{i\phi_i\left(\vec{s}+\vec{\delta}_i\right)}\,.
\end{equation}
Since the fiber center is not necessarily aligned perfectly with the image center, a small offset $\vec{\delta}_i$ can arise between the zero-coordinate of the phase and amplitude maps and the sky intensity distribution.

The position integral in the numerator of Eq.~(\ref{eq:appx-response-full}) can be computed very efficiently with the fast fourier transform (FFT) algorithm. However, the grid of rectangular frequencies does not necessarily align with the $\left(u,v\right)$-coordinates of the measurements. A common technique in radio and optical interferometry, known as gridding, is to interpolate the complex visibilities between the regular grid points of the FFT \citep{gridding-article, 1999ASPC..180..127B}.

To implement the phase and amplitude maps numerically, we produced a complex screen for each telescope, following the procedure in \cite{2021A&A...647A..59G}. The maps already account for the instrument's aperture and are smoothed with a $\SI{10}{mas}$ Gaussian kernel that models residual tip-tilt jitter from the AO system \citep{2019A&A...625A..48P}. The resulting complex screen has the same dimensions as the actual image, and both are multiplied pixel-wise before FFT and gridding are performed\footnotemark. Because the aberrations differ between telescopes, we need to perform the computation separately on each baseline.
\footnotetext{We use the ducc0 library for FFT and gridding \citep{2021A&A...646A..58A} which provides C++17 code with a comprehensive Python interface, cf. \object{https://gitlab.mpcdf.mpg.de/mtr/ducc}}

Taking into account the temporal and spectral variability naively would lead to a slightly different image for each exposure and each spectral channel and further curtail any computational gain achievable by gridding. To avoid this, we write the overall sky brightness distribution as
\begin{align}
I\left(\vec{s},\lambda, t, p\right) =& f_{\nu_1}\,\left(\lambda\right) I_\mathrm{SgrA}\left(t,p\right)\delta\left(\vec{s}-\vec{s}_\mathrm{SgrA}\right)\nonumber\\
+& f_{\nu_2}\,\left(\lambda\right)\left[I_\mathrm{Img}\left(\vec{s}\right) + \sum_{j=1}^{N_\mathrm{PS}} I_j\, \delta\left(\vec{s}-\vec{s}_j\right)\right]\,,
\label{eq:appx-response-decomposition}
\end{align}
where the first line corresponds to Sgr~A*, a variable point source with spectral index $\nu_1$ and the second line represents the image of faint sources plus $N_\mathrm{PS}$ additional point sources whose spectral index is $\nu_2$. The spectral distributions are given by
\begin{equation}
f_\nu\left(\lambda\right) = \left(\frac{\lambda}{\SI{2.2}{\mu m}}\right)^{-\nu-1}\,,
\label{eq:appx-response-spectral}
\end{equation}
where we choose the reference wavelength at the center of the K-band observed by GRAVITY. In this context, the spectral indices correspond to the observed spectrum, namely, the intrinsic spectrum of the sources altered by interstellar absorption and reddening, the Earth's atmosphere, and instrumental transmission. However, the complex visibilities in Eq.~(\ref{eq:appx-response-full}) are primarily sensitive to the difference $\nu_1-\nu_2$, rather than to the individual spectral indices due to the normalization by the total flux.

We then approximate the bandpass integration by an average over $N_\lambda$ points, distributed linearly over each channel such that the numerator of Eq.~(\ref{eq:appx-response-full}) becomes
\begin{align}
\frac{1}{N_\lambda} \sum_{i=1}^{N_\lambda} P_{\lambda_0}\left(\lambda_i\right) \left\{ f_{\nu_1}\left(\lambda_i\right)\, I_\mathrm{SgrA}\left(t,\,p\right)\,\ee^{-2\uppi i \vec{b}\cdot\vec{s}_\mathrm{SgrA}/\lambda_i} + \right.\nonumber\\
\left.
f_{\nu_2}\left[\sum_{j=1}^{N_\mathrm{PS}} I_j\,\ee^{-2\uppi i \vec{b}\cdot\vec{s}_j/\lambda_i} + \int \dd\vec{s}\, I_\mathrm{Img}\left(\vec{s}\right)\,\ee^{-2\uppi i \vec{b}\cdot\vec{s}/\lambda_i} \right]
 \right\}\,.
 \label{eq:appx-response-nominator}
\end{align}
Because we implemented the last integral as an FFT and a subsequent gridding operation, evaluating it for a fine-gridded set of $\lambda$ values is computationally not very expensive as long as the image inside the Fourier transform is mono-chromatic. The computational steps to evaluate Eq.~(\ref{eq:appx-response-nominator}) are summarized in the following. 
\begin{enumerate}
\item We use a FFT followed by gridding to compute
\begin{equation}
\int\dd\vec{s}~A_i\left(\vec{s}\right)A_j\left(\vec{s}\right) I_\mathrm{Img}\left(\vec{s}\right) \ee^{-2\uppi i \vec{b}\cdot\vec{s}/\lambda + i\phi_i\left(\vec{s}\right)-i\phi_j\left(\vec{s}\right)}\,
\end{equation}
simultaneously for all exposures, all spectral channels, and all fine-grained $\lambda$ values within a channel.
\item For all additional point sources, we obtain $\ee^{-2\uppi i \vec{b}\cdot\vec{s}/\lambda}$ at each exposure, channel and sub-resolution in $\lambda$ and multiply it by the phase- and amplitude-maps at the position of the source. Hereby, we approximate the aberration maps at the sources' actual positions by the aberration maps at the center of their position prior. Since the maps are very flat on the scale of the position uncertainty, the error introduced by this approximation is small, however it significantly simplifies the computation.
\item We multiply the expression for the time-dependent point source by an exposure- and polarization-dependent brightness $I_\mathrm{SgrA}\left(t, p\right)$ and all further point sources by their constant brightness $I_j$ which is independent of the spectral channel.
\item We then account for the spectral distribution by multiplying each source at each $\lambda$-coordinate with the appropriate value from Eq.~(\ref{eq:appx-response-spectral}). This multiplication is the same for all exposures.
\item Finally, we add up all components, multiply by the spectral bandpass and perform the wavelength integration or bandwidth smearing as an average over all fine-grained $\lambda$-steps within one channel.
\item The procedure is repeated independently for all six GRAVITY baselines due to the changing aberration maps between baselines.
\end{enumerate}
This implementation is very flexible in that more complicated spectra than the power-law of Eq.~(\ref{eq:appx-response-spectral}) can easily be accommodated. It can also readily be generalized to accomodate additional spectral components and different bandpass shapes.

We use calibration measurements of the GRAVITY spectral transmission to investigate the effect of bandwidth smearing and to compare the full measured bandpass to a top hat approximation. We find that the top-hat affects the computed visibilities on a level much smaller than the typical GRAVITY error bars and, therefore, we chose the approximation for this work. To determine the required sub-sampling steps per spectral channel, we iteratively increase $N_\lambda$ until the visibilities have converged to a stable value. This is the case by a good margin at $N_\lambda = 100$.

To implement the denominator term of Eq.~(\ref{eq:appx-response-full}), we note that the spectral and positional integration can be decoupled under the parametrization in Eq.~(\ref{eq:appx-response-decomposition}). Furthermore, assuming a top-hat bandpass with center $\lambda_0$ and width $2\delta\lambda$:
\begin{equation}
P\left(\lambda\right) = \mathrm{rect}\left(\frac{\lambda-\lambda_0}{2\delta\lambda}\right) = \begin{cases}
1   & \mathrm{if}  \left|\lambda-\lambda_0\right| < 0\\
1/2 & \mathrm{if} \left|\lambda-\lambda_0\right| = 0\\
0   & \mathrm{otherwise,}
\end{cases}\,,
\end{equation}
the wavelength integral is evaluated analytically to:
\begin{equation}
\int \dd\lambda P\left(\lambda\right)\,f_\nu\left(\lambda\right) = \frac{2.2\,\upmu\mathrm{m}}{\nu}\left[\left(\frac{\lambda_0-\delta\lambda}{2.2\,\upmu\mathrm{m}}\right)^{-\nu} - \left(\frac{\lambda_0+\delta\lambda}{2.2\,\upmu\mathrm{m}}\right)^{-\nu}\right]\,.
\label{eq:appx-response-wavelengthintegral}
\end{equation}

The integral in the spatial directions, on the other hand, can be computed readily as a sum over all image pixels multiplied by the squared amplitude maps. To this, we add the intensity of potential extra point sources multiplied by the fiber damping. For the variable component, the spatial integral is simply given by the product between light curve and fiber damping. Either contribution is then multiplied by Eq.~(\ref{eq:appx-response-wavelengthintegral}), then their summation gives the denominator of Eq.~(\ref{eq:appx-response-full}).

Once we have evaluated the deterministic part of the response function, we multiply it by the time- and baseline dependent amplitude calibration factor, introduced in Sect.~\ref{sec:method-selfcal}. 

With this setup, one full response evaluation, including all exposures, baselines, and channels of a typical observing night with 21 exposures takes \SI{0.4}{\s} on a single laptop CPU (Intel i7 \SI{1.90}{GHz}).

\section{Formal description of the inference scheme}
\label{appx:model}
Given the sky model from Sect.~\ref{sec:method-prior} (cf. also Eq.~\ref{eq:appx-response-decomposition}) and the self-calibration approach for the visibility amplitudes described in Sect.~\ref{sec:method-selfcal}, the full set of model parameters to be inferred is
\begin{align}
\theta = \left\{I_\mathrm{Img}\left(\vec{s}\right),\, I_\mathrm{SgrA}\left(t,p\right),\, \vec{s}_\mathrm{SgrA},\, I_1, ... I_\mathrm{NP},\, \vec{s}_1,\, ... \right.\nonumber\\ 
\left.
...\vec{s}_\mathrm{NP},\, \nu_1,\,\nu_2,\, C\left(t,\,b\right) \right\}\,.
\end{align}
For a clearer notation, we have written the spatial ($\vec{s}$), temporal ($t$), baseline ($b$) and the polarization ($p$) dependence of these quantities as functional arguments, but we use them, in particular, to label the discretized coordinates. The degrees of freedom inherent to each of these parameters are counted in Eq.~(\ref{eq:method-dof}). The prior model for the image of faint sources is provided in Eq.~(\ref{eq:method-faint-sky-prior}), and for the remaining parameters we have:
\begin{align}
\ln I_\mathrm{SgrA}\left(t,p\right) &\hookleftarrow \prod_{j=1}^{N_\mathrm{exp}} \prod_{p=1,2} \mathcal{G}\left[\ln \left.I_\mathrm{SgrA}\left(t_j,p\right)\, \right| 0, 1\right]\,,
\label{eq:appx-inference-prior1}\\
\vec{s}_\mathrm{SgrA} &\hookleftarrow \mathcal{G}\left(\left.\vec{s}_\mathrm{SgrA}\,\right| \vec{\mu}_\mathrm{SgrA},\, \sigma_\mathrm{SgrA}\mathbb{1}\right)\,,
\\
\ln I_i &\hookleftarrow \mathcal{G}\left(\left.\ln I_i \right|\, 0, 1\right) \quad \mathrm{for}\quad i=1, ...\, N_\mathrm{PS}\,,
\\
\vec{s}_i &\hookleftarrow \mathcal{G}\left(\left.\vec{s}_i\,\right| \vec{\mu}_i,\, \sigma_i\mathbb{1}\right) \quad \mathrm{for}\quad i=1, ...\, N_\mathrm{PS}\,,
\\
\nu_i &\hookleftarrow \mathcal{G}\left(\left.\nu_i\right|\, \mu_\mathrm{\nu_i},\, \sigma_\mathrm{\nu_i}\right) \quad \mathrm{for} \quad i=1,2\,,
\\
C_i\left(t,\,b\right) &\hookleftarrow \prod_{b=1}^{6} \prod_{j=1}^{N_\mathrm{exp}} \mathcal{G}\left[ \left.C_i\left(t_j\right)\right|\, 1,\, \sigma_C\right] \,,
\label{eq:appx-inference-prior2}
\end{align}
where the ''$\hookleftarrow$" indicates the probability density function from which a variable is drawn a priori and $N_\mathrm{exp}$ counts the exposures in the data set. MGVI works with standardized coordinates, that is, all the model parameters are mapped to a set of auxiliary parameters $\xi,$ which follow a standard normal distribution. This mapping is obtained by a two-step process,
\begin{equation}
\theta = \left[\mathcal{F}^{-1}_{\mathcal{P}\left(\theta\right)} \circ \mathcal{F}_{\mathcal{G}\left(\xi|0,\mathbb{1}\right)}\right]
\left(\xi\right)\,,
\label{eq:appx-signal-operator}
\end{equation}
where $\mathcal{F}$ and $\mathcal{F}^{-1}$ denote the cumulative probability density function and its inverse, respectively. The first step translates a standard normal distributed to an uniformly distributed variable. In the second step, this uniform distribution is transformed to the prior specified in Eq.~(\ref{eq:method-faint-sky-prior}) and Eqs.~(\ref{eq:appx-inference-prior1}) to (\ref{eq:appx-inference-prior2}). 

For brevity, we introduce an operator notation in which $\mathbf{\mathcal{S}}$ is the signal operator and corresponds to the application of Eq.~(\ref{eq:appx-signal-operator}). The response operator, $\mathbf{\mathcal{R}}$, computes complex visibilities for any realization of $\theta$ as specified by Eq.~(\ref{eq:appx-response-full}). Finally, the complex visibilities are transformed to visibility amplitudes and closure phases by the $\mathbf{\mathcal{A}}$ and $\mathbf{\mathcal{C}}$ operator, respectively.

In this notation, the negative logarithm of the likelihood discussed in Sect.~\ref{sec:method-likelihood} is expressed as:
\begin{align}
&-\ln \mathcal{P}\left(\vec{d}|\xi\right)\, \widehat{=}\,
\frac{1}{2}\left[\ee^{i\left(\mathbf{\mathcal{C}}\circ \mathbf{\mathcal{R}}\circ\mathbf{\mathcal{S}}\right)\left(\xi\right)} -\ee^{i\mathbf{\mathcal{C}}\left(\vec{d}\right)}\right] N_\Phi^{-1} \left[\ee^{i\left(\mathbf{\mathcal{C}}\circ \mathbf{\mathcal{R}}\circ\mathbf{\mathcal{S}}\right)\left(\xi\right)} -\ee^{i\mathbf{\mathcal{C}}\left(\vec{d}\right)}\right]^\dagger
\nonumber\\ 
&+\frac{1}{2}\left[\left(\mathbf{\mathcal{A}}\circ \mathbf{\mathcal{R}}\circ\mathbf{\mathcal{S}}\right)\left(\xi\right) -\mathbf{\mathcal{A}}\left(\vec{d}\right)\right] N_\rho^{-1} \left[\left(\mathbf{\mathcal{A}}\circ \mathbf{\mathcal{R}}\circ\mathbf{\mathcal{S}}\right)\left(\xi\right) -\mathbf{\mathcal{A}}\left(\vec{d}\right)\right]^\dagger\,,
\nonumber\\
\end{align}
where ``$\widehat{=}$'' denotes equality up to addition of a constant and $N_\Phi^{-1}$ and $N_\rho^{-1}$ are the diagonal covariance matrices of the closure phases and visibility amplitudes (see Sect.~\ref{sec:method-likelihood}). In the first line, we approximate the closure phases by their position on the unit circle, which mitigates the problem of phase wraps. This represents an approximation to the full likelihood, which becomes precise in the limit of small differences and down-weights outliers. The corresponding negative log-prior, on the other hand, is now simply given by
\begin{equation}
-\ln \mathcal{P}\left(\xi\right) ~\widehat{=}~ \frac{1}{2} \xi \mathbb{1} \xi^\dagger\,.
\end{equation}

In the inference, MGVI approximates the full posterior by a multivariate Gaussian distribution with covariance $\Xi \simeq I\left(\xi\right)^{-1}$, where $I\left(\xi\right)$ is a generalization of the Fisher metric defined as
\begin{align}
I\left(\xi\right) &= \mathbb{1} + \left[\frac{\partial\, \ee^{i\left(\mathbf{\mathcal{C}}\circ \mathbf{\mathcal{R}}\circ\mathbf{\mathcal{S}}\right)\left(\xi\right)}}{\partial\xi}\right]
N_\Phi^{-1}
\left[\frac{\partial\, \ee^{i\left(\mathbf{\mathcal{C}}\circ \mathbf{\mathcal{R}}\circ\mathbf{\mathcal{S}}\right)\left(\xi\right)}}{\partial\xi}\right]^\dagger
\nonumber\\
&+\left[\frac{\partial\, \left(\mathbf{\mathcal{A}}\circ \mathbf{\mathcal{R}}\circ\mathbf{\mathcal{S}}\right)\left(\xi\right)}
{\partial\xi}\right]
N_\rho^{-1}
\left[\frac{\partial\, \left(\mathbf{\mathcal{A}}\circ \mathbf{\mathcal{R}}\circ\mathbf{\mathcal{S}}\right)\left(\xi\right)}
{\partial\xi}\right]^\dagger\,.
\label{appx:inference-fisher-matrix}
\end{align}
Here, the first term originates from the prior and the latter two from the likelihood. The NIFTy package \citep{2019ascl.soft03008A}, which we have used to implement the inference, facilitates auto-differentiation, such that only the derivatives of individual operators need to be implemented and an operator sequence can be differentiated automatically by applying the chain rule.

\section{Hyper-parameters of the inference}
\label{appx:hyperparameters}

\begin{table}
\caption{Iteration scheme for the posterior exploration with MGVI.}
\label{tab:appx-hyperparameters-scheme}
\centering
\begin{tabular}{c c c}
iteration step & \# minimization steps & \# samples\\
\hline
1 - 3 & 4 & 3\\
4 \& 5 & 5 & 5\\
7 \& 8 & 8 & 6\\
9 - 12 & 10 & 8\\
13 - 20 & 15 & 12\\
21 - 25 & 20 & 15\\
26 - 30 & 30 & 20
\end{tabular}
\end{table}

The performance of MGVI depends on a series of hyper-parameters which need to be set by the user. Among them is the number of MGVI iterations and the number of samples and minimization steps at each iteration which we jointly refer to as the iteration scheme. In addition, it is possible to choose between different minimizers and iteration controllers for the minimization of the KL and for drawing samples at a given position.

We determined these hyper-parameters in a two step procedure and first consider a mock data set, for which we took a representative $\left(u,v\right)$-coverage from the 2019 observing series. We then specified a model with a variable Sgr~A* at the center, two faint sources in its vicinity and a brighter S2-like source to the north east. The model provides a prediction for closure phases and visibility amplitudes, to which we added a random Gaussian noise realization.

In the analysis of the mock data set, we varied all the aforementioned hyper-parameters with the aim to optimally recover the ground truth. We find that an iteration controller that checks the relative energy change between steps works best for drawing samples and minimizing the KL, and for the latter task we picked a Newton Conjugate Gradient minimizer. None of our results show significant changes after 30 MGVI iterations. 

\cite{2020arXiv200205218A} noted that tempering during the early MGVI iterations could improve their results and to this end, they iterated between minimizing the closure and amplitude likelihoods separately. We do not find a similar improvement in our reconstructions and thus consider the full likelihood at all MGVI steps.

The performance on the actual data can differ from the mock-imaging tests, for example, if not all noise correlations are modeled or if some residual systematic effects remain unaccounted for by the response function. Therefore, in a second step, we revised our iteration scheme now considering actual observations. This resulted in an overall decrease in the number of minimization steps and an increase in the number of samples at each MGVI iteration. Accordingly, MGVI reaches the minimum more slowly and overfitting is impeded. We can thus view this revision as the adoption of a less aggressive and more conservative approach. 

We summarize the minimization scheme in Table~\ref{tab:appx-hyperparameters-scheme}. The consistency of our images across multiple nights and observing periods (cf. Sect.~\ref{sec:results-s300-detection}) and injection tests (cf. Sect.~\ref{sec:results-injection}) verify that our result is not an artifact of the minimization routine.

\section{Identification of stars}
\label{sec:appx-identification}

\begin{figure*}[t!]
        \centering
        \includegraphics[width=17.5cm]{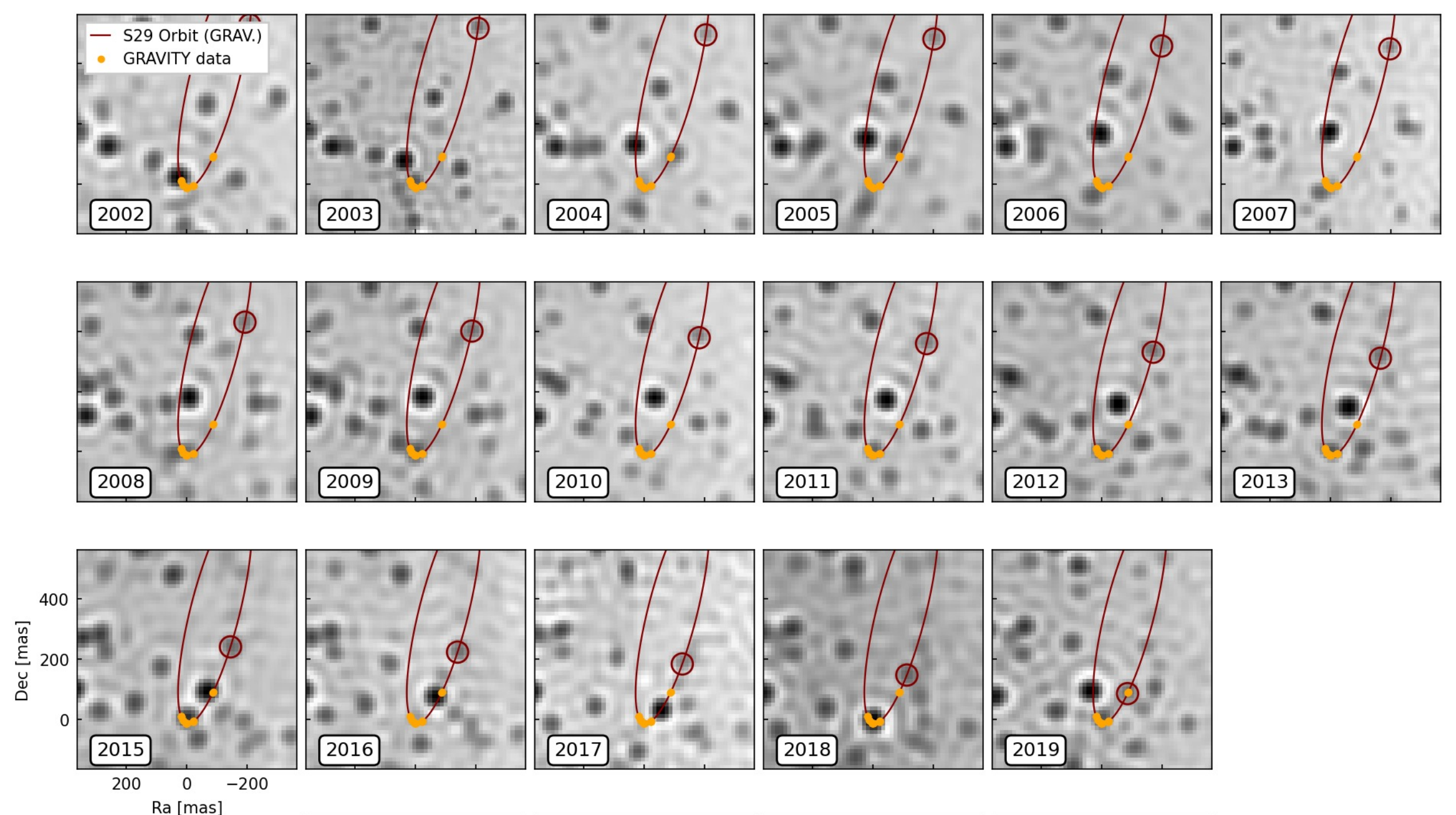}
        \caption{Compilation of deconvolved NACO images cut to the central $\pm 400$,mas. The small circle in dark red marks the position of S29  on its orbit (dark red ellipse) in each panel. The orange data points are the 2019 and 2021 GRAVITY measurements. Evidently, S29 can be traced consistently over two decades. We note that around 2015 - 2017 the star is confused with other stars, such that the astrometry becomes unreliable, but the identification remains unambiguous.}
        \label{fig10}
\end{figure*}

\begin{figure}
\includegraphics[width=6cm]{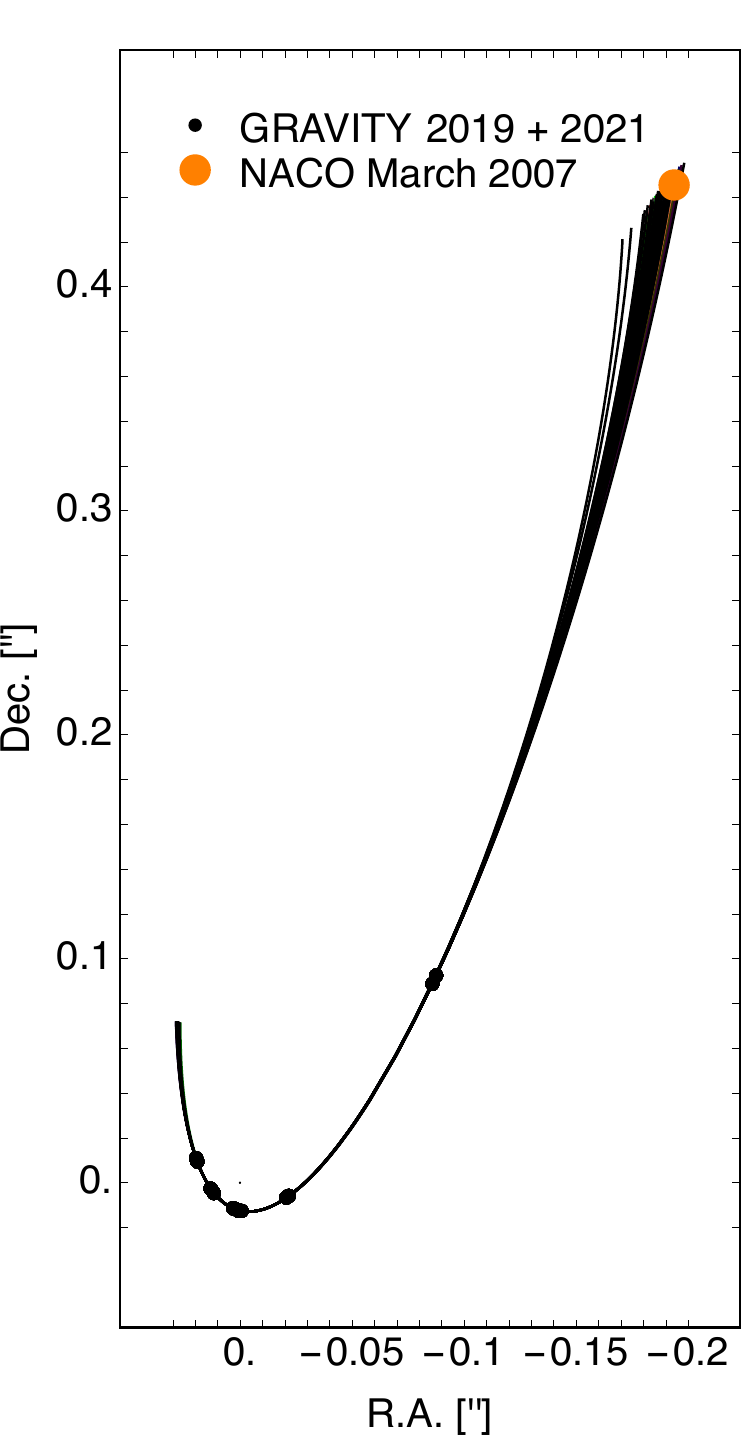}
\caption{Backwards prediction of the GRAVITY data for $\mathrm{S29}_\mathrm{GRAVITY}$ leads to a position in March 2007 compatible with the position of $\mathrm{S29}_\mathrm{NACO}$. The black points show the GRAVITY data from 2021 and 2019. The set of black orbits is obtained by bootstrapping these data and fitting each of these mock data sets. The family of orbits naturally matches the position obtained from NACO (orange disk). The epoch March 2007 was chosen since it is the one used in \cite{2009ApJ...692.1075G} to define the names of the S-stars.}
\label{fig:appx-s29-backward}
\end{figure}

The GRAVITY images illustrate how S29 flies through pericenter in 2021 and here we comment on the crossidentification of the star with the earlier, AO-based NACO data. For that purpose, we differentiate between $\mathrm{S29}_\mathrm{GRAVITY}$ (i.e., the star labeled S29 throughout this publication and observed with GRAVITY) and $\mathrm{S29}_\mathrm{NACO}$, the star label S29 in \citet{2009ApJ...692.1075G}. 

\subsection{Identifying $\mathrm{S29}_\mathrm{GRAVITY}$ with $\mathrm{S29}_\mathrm{NACO}$}

In Fig.~\ref{fig10}, we show a compilation of deconvolved NACO images from 2002 to 2019, which allow us to trace $\mathrm{S29}_\mathrm{NACO}$ through the years. In 2019, the star was observed with NACO and GRAVITY, and we obtain matching positions for $\mathrm{S29}_\mathrm{NACO}$ and $\mathrm{S29}_\mathrm{GRAVITY}$. 

Using only the 2021 GRAVITY data for $\mathrm{S29}_\mathrm{GRAVITY}$, we can fit an orbit and extrapolate back to 2019, yielding a position consistent with the 2019 astrometric points of GRAVITY for $\mathrm{S29}_\mathrm{GRAVITY}$. Furthermore, the orbit from the full interfermetric data set, that is, GRAVITY astrometric positions from 2021 to 2019, reliably predicts the $\mathrm{S29}_\mathrm{NACO}$ position in earlier AO epochs. In particular, it agrees with the 2007 reference map in \cite{2009ApJ...692.1075G}, as demonstrated in Fig.~\ref{fig:appx-s29-backward}. 

We conclude that our identification of $\mathrm{S29}_\mathrm{GRAVITY}$ with $\mathrm{S29}_\mathrm{NACO}$ is robust and can be done both forwards from the AO epochs to GRAVITY as well as backwards from the GRAVITY data to the NACO images obtained $15-20$ years earlier.

\subsection{Excluding other identifications for $\mathrm{S29}_\mathrm{GRAVITY}$}

Conserved dynamical quantities can be used to tag stars, since the conserved quantity needs to apply at each point on the orbit for a given star. Here, we use the $z$-component of the angular momentum vector, $h_z = x v_y - y v_x$, which can be derived from astrometric data alone. For  $\mathrm{S29}_\mathrm{GRAVITY}$, we can determine that value for four occasions in 2021: during each of the four observing campaigns, we were not only able to determine the positions, but also the proper motions. This yields $h_z=(-1.09{\pm}0.03)\times10^{20}~\mathrm{m}^2/\mathrm{s}$, where the value is the mean over the four epochs and the error the corresponding standard deviation. The two 2019 points for $\mathrm{S29}_\mathrm{GRAVITY}$ yield $h_z \approx -1.1 \times 10^{20} \mathrm{m}^2/\mathrm{s}$, consistent with the 2021 value.

This excludes, for example, that $\mathrm{S29}_\mathrm{GRAVITY}$ is identical to the source candidate from  \cite{2021ApJ...918...25P}, which we henceforth refer to as $\mathrm{S62}_\mathrm{Peissker}$. The latter star has a significantly smaller z-component for the angular momentum vector, $h_z\approx+1.4 \times 10^{19} \mathrm{m}^2/\mathrm{s}$. Indeed, a star with such a high value of $h_z$ as $\mathrm{S29}_\mathrm{GRAVITY}$ would need to show an (tangential) on-sky motion of $\approx 880\,$km/s, for a projected radius of $0.1\,$mas, at appocenter.

We also inspected the two 2019 data sets obtained on $\mathrm{S29}_\mathrm{GRAVITY}$. Using the same imaging technique as in Sect.~\ref{sec:discussion-clean} and \cite{2021A&A...645A.127G}, we find that there is only one dominant source in the field. Furthermore, the residual level is fainter than $m_K = 19$. The structure of these residuals can be attributed to S2, which is significantly brighter than $\mathrm{S29}_\mathrm{GRAVITY}$ but located further away from the field center and thus appears significantly damped by the Gaussian acceptance profile of the GRAVITY fibers.

\subsection{Looking for the counterpart of $\mathrm{S62}_\mathrm{Peissker}$}

There are two sets of GRAVITY pointings which contain the position predicted by the $\mathrm{S62}_\mathrm{Peissker}$ orbit in \cite{2020ApJ...889...61P}. First, we would have expected it to show up in the 2019 GRAVITY pointings to $\mathrm{S29}_\mathrm{GRAVITY}$. Second, we pointed to the north-west of Sgr~A* this year as part of the mosaicing data set (see Table~\ref{tab:data-observations-widefield} and Fig.~\ref{fig:res-offpointings-summary}). In the former, we detect S29 at the expected position and with the expected velocity, as discussed in the previous sub-section. In the images from the latter data set (see Fig.~\ref{fig:res-offpointings-summary}), we can identify S42, S2 and Sgr~A*. However, we do not find any further sources. Given the predicted brightness of $\mathrm{S62}_\mathrm{Peissker}$, it would have to be outside the FOV of either pointing. In 2022, the star is predicted even closer to Sgr~A* and should then appear in the central pointing, that is, when positioning the GRAVITY fibers directly onto Sgr~A*.

\section{Probabilistic view of the imaging results}
\label{appx:probabilistic-results}

\begin{figure*}
        \sidecaption
        \includegraphics[]{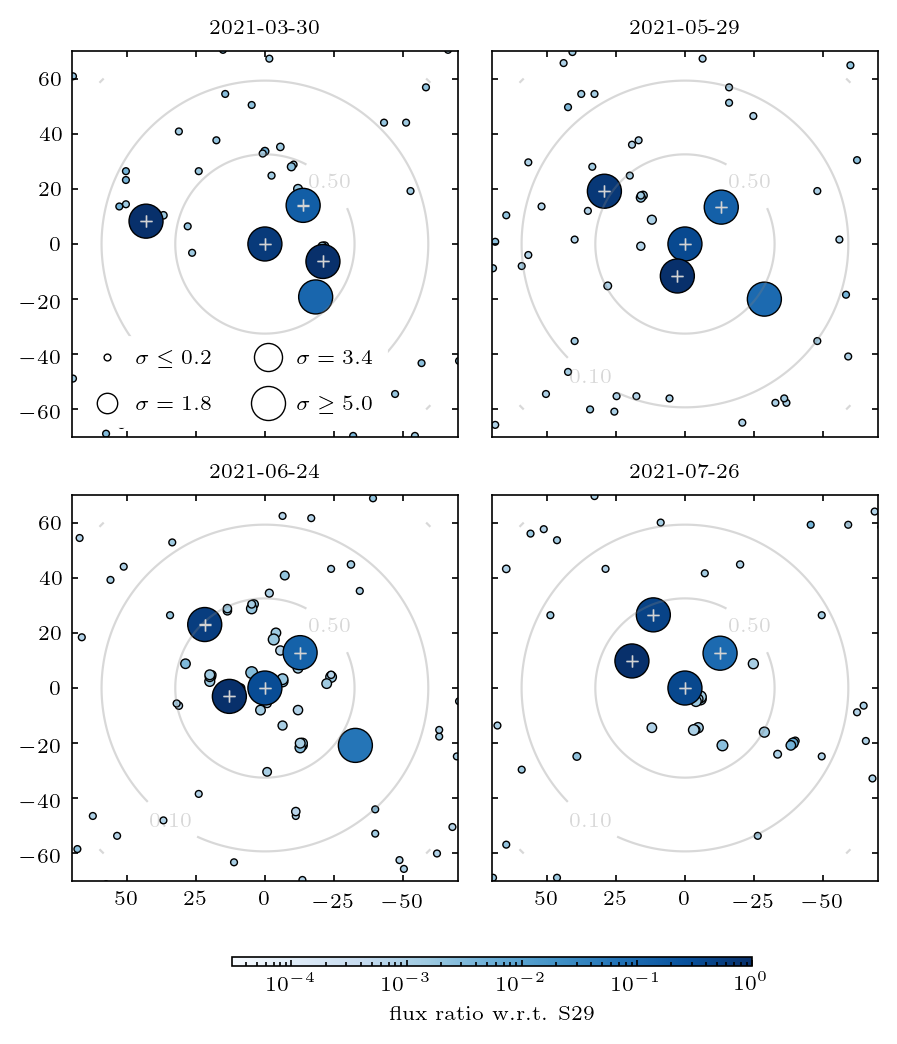}
        \caption{Statistical view of the imaging results shown in Fig.~\ref{fig:results-images}. The symbol color indicates the flux of a source candidate normalized to S29, while its size represents the significance (see Appendix~\ref{appx:probabilistic-results} for details). The Sgr A* flux depicted here has to be multiplied by the light curve at each exposure to arrive at the true flux ratios. For stars which are modeled as a point source, the position uncertainty is indicated in gray.}
        \label{fig:appx-epochwise-sources}
\end{figure*}

\begin{figure*}
        \centering
        \includegraphics[]{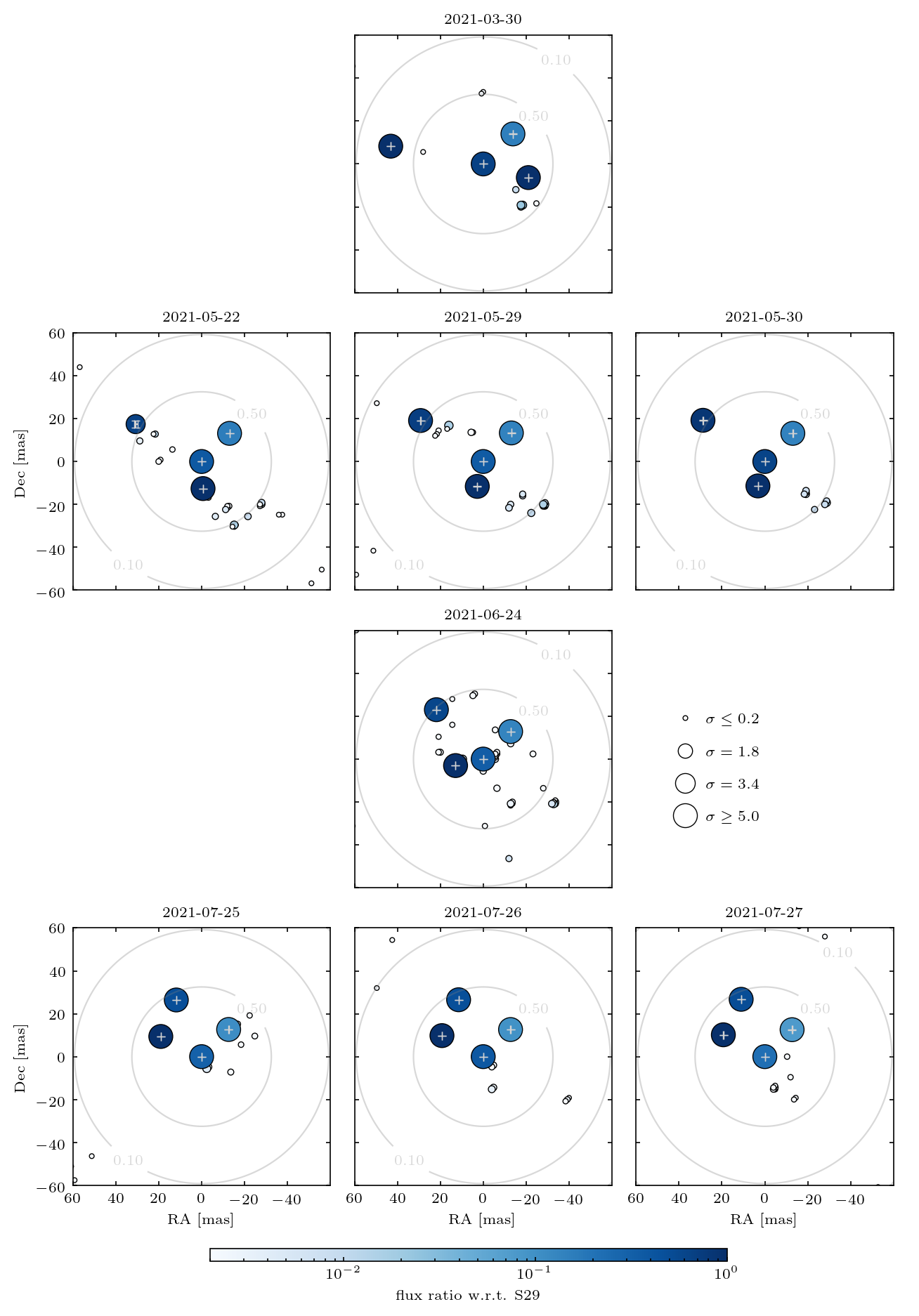}
        \caption{Summary of all imaging runs for Sgr~A*-centered exposures. Each panel combines the samples from ten imaging runs with independent random seeds (see Appendix~\ref{appx:probabilistic-results} for details). The symbol color indicates the flux of a source candidate normalized to S29, while its size represents the significance. The Sgr A* flux depicted here has to be multiplied by the light curve at each exposure to arrive at the true flux ratios. For stars modeled as a point source, the position uncertainty is indicated in gray.}
        \label{fig:appx-results-summary}
\end{figure*}

\begin{figure*}
        \sidecaption
        \includegraphics[]{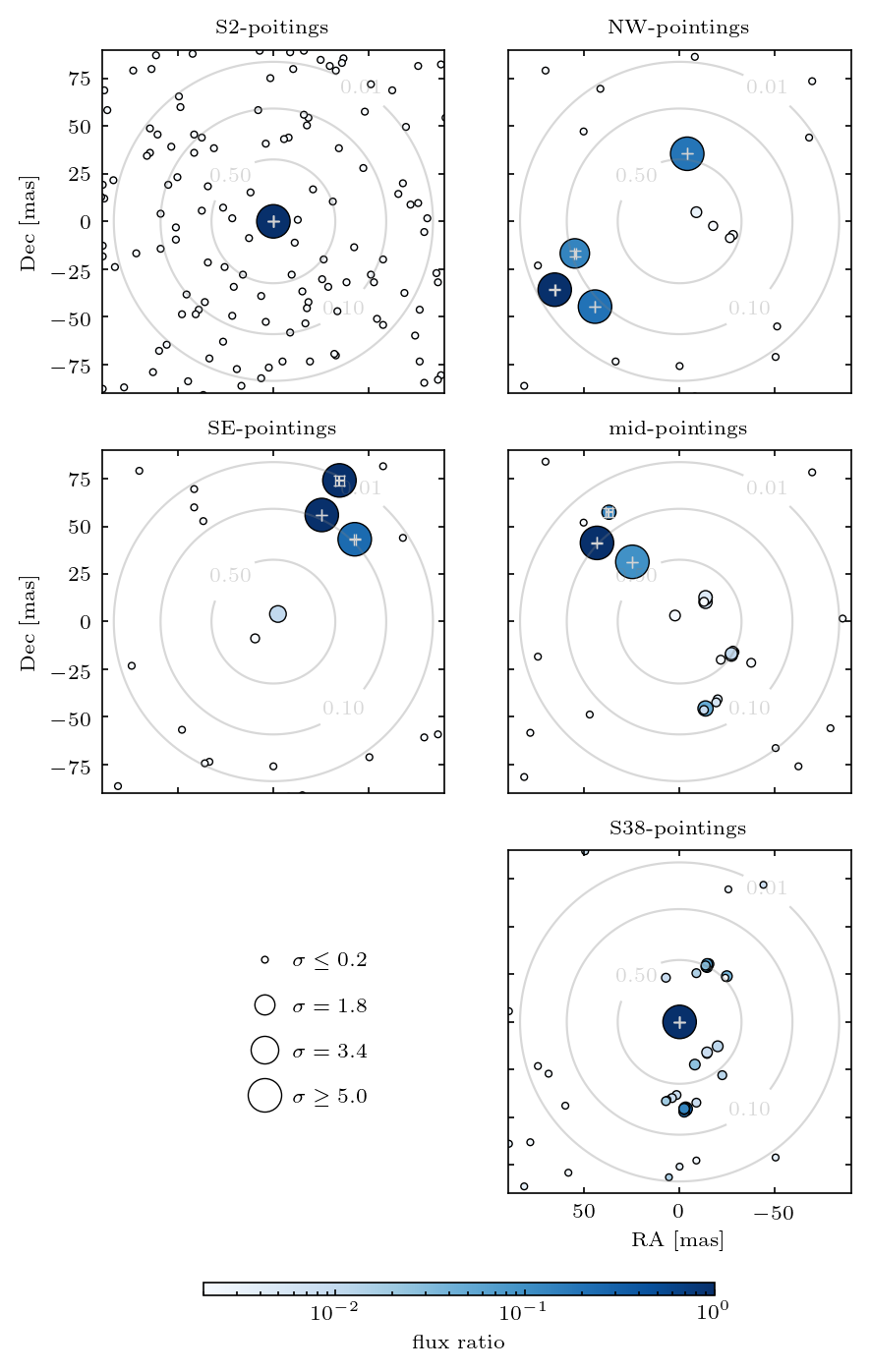}
        \caption{Summary all imaging runs for the mosaicing data set. Each panel combines the samples from ten imaging runs with independent random seeds (see Appendix~\ref{appx:probabilistic-results} for details). The symbol color indicates the flux of a source candidate, while its size represents the significance. The flux in the images is normalized to S2 in the top left panel, to S38 in the bottom right panel, and to S29 in all other instances. The Sgr A* flux depicted here has to be multiplied by the light curve at each exposure to arrive at the true flux ratios. For stars modeled as a point source, the position uncertainty is indicated in gray.}
        \label{fig:appx-results-offpointings-summary}
\end{figure*}

In Figs.~\ref{fig:results-images} and \ref{fig:results-offpointing-images}, we show the main imaging results of this publication. These images are computed as the mean over all samples in a single \gr~run and have been convolved with a Gaussian of $1.6$ standard deviation to account for the typical size of the CLEAN beam. 

However, the MGVI samples contain information beyond the mean and can also be used to estimate the uncertainty of the result. Since our sky model for the GC is a collection of point sources and the main goal of the analysis is the search for faint, yet unknown stars, we visualize the information contained in the MGVI samples in the following way. In the mean image, we select all pixels whose flux exceeds the background by at least a factor of 10. This threshold is chosen to be below the optimal sensitivity found in Sect.~\ref{sec:results-injection}. Then, for all potential sources identified in this way, we compute the flux variance in the corresponding pixel and express the mean flux in units of its standard deviation. We summarize the position of all potential sources in a RA-Dec plot, as the one shown in Fig.~\ref{fig:appx-epochwise-sources}. Thereby, the symbol size indicates the mean flux in units of its standard deviation, that is, the significance of the source candidate, while the color signals its flux. For known sources with a Gaussian position prior, \gr~directly infers the coordinates and the standard deviation of this estimate is overdrawn on the source symbol. 

This statistical view of the imaging results is provided in Fig.~\ref{fig:appx-epochwise-sources} for the same \gr~runs, which are shown in Fig.~\ref{fig:results-images}. As Fig.~\ref{fig:appx-epochwise-sources} demonstrates, S300 corresponds not only to the brightest pixel in the image of faint sources, but it is also the only pixel where flux is detected at a high significance. The only exception  here is the July result, where S300 is most strongly affected by fiber damping and \gr~fails to detect its flux in a single high-significance pixel. To evaluate the \gr~results, we produce and inspect this type of figure for each of the ten \gr~imaging runs separately. The results of this inspection are reported in Sect.~\ref{sec:results-s300-detection}.

Further, we also combine the samples from all ten \gr~runs into a single figure for a joint representation. These are shown, for all nights considered in Tables~\ref{tab:data-observations-sgra} and~\ref{tab:data-observations-widefield}, in Figs.~\ref{fig:appx-results-summary} and \ref{fig:appx-results-offpointings-summary} respectively. Here, we did not perform a selection of runs, which means that the figures also contain poorly converged runs and runs that failed to identify S300. Furthermore, as we report in Sect.~\ref{sec:results-s300-detection}, the position at which S300 is detected can vary by a single pixel. The same holds for sources in the mosaicing data set, which are not modeled as a point source which is superimposed on the image. Both effects, the inclusion of all runs and the position uncertainty, lead to a decrease in the significance estimate of faint sources in the joint representation. The effect is very apparent from the comparison of Fig.~\ref{fig:appx-epochwise-sources} and Fig.~\ref{fig:appx-results-summary}. Also for this reason, the evaluation of the \gr~results run-by-run as presented in Sect.~\ref{sec:results-s300-detection} is a very important step of the analysis.

\section{Source injection tests}
\label{appx:injection-test}

\begin{figure*}
        \sidecaption
        \includegraphics[]{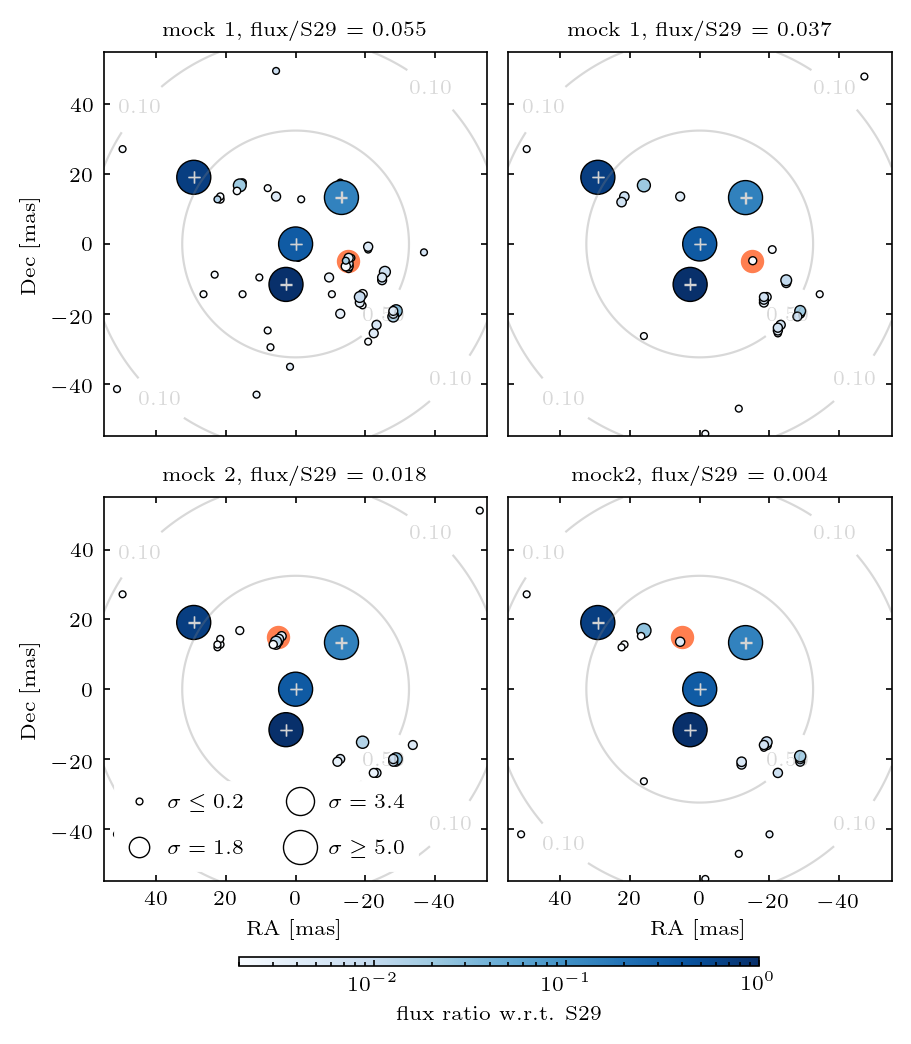}
        \caption{Results of injection tests for the May 29 data set. The position of the injected source is at $(-15.0, -5.0)\,\si{mas}$ from Sgr~A* for mock 1, at $(5.0, 15.0)\,\si{mas}$ for mock 2 and marked by an orange circle in the respective plots. We show the combined samples from ten imaging runs with varying random seed and indicate the mean inferred flux, normalized by S29, by symbol color and the significance of a source candidate by the size of its symbol (see Appendix~\ref{appx:probabilistic-results} for details). In this fashion of display, variation of a faint source's position by a pixel between the imaging runs leads to a decrease in the estimated significance, even if the detection is very robust ($> 5\sigma$) for individual runs. The Sgr A* flux depicted here has to be multiplied by the light curve at each exposure to arrive at the true flux ratios. For stars that are modeled as a point source, the position uncertainty is indicated in gray.}
        \label{fig:results-injection}
\end{figure*}

We performed a series of injection tests to further corroborate the result from Sect.~\ref{sec:results-s300-detection}. Injecting a source into the data and trying to recover it in the images has the additional benefit of making it possible to dial down the source's flux step-by-step and in this fashion test our sensitivity.

To create the data for the injection tests, we pick the imaging run with the cleanest result from the 2021-05-29 night, this corresponds to the image shown in Fig.~\ref{fig:results-images}. The posterior mean gives a model prediction for closure phases and visibility amplitudes, which we subtract from the data to obtain the residuals. We then augment the model by an additional faint source, again compute closures and amplitudes, and, finally, we add back the residuals. In comparison to creating mock data, where we would specify the model by hand and then add a random Gaussian noise realization, this approach preserves the effect of any instrumental systematics as well as correlations between individual data points, which are unaccounted for by the prior model and the likelihood.

In summary, we tested mock sources at two different positions, each with four values for its flux. The first position is at $\vec{s}_\mathrm{mock1} = (-15.0, 5.0)\,\si{mas}$ from Sgr~A* and thus relatively close to S300. The second location we pick at identical separation from the field center with $\vec{s}_\mathrm{mock2} = (5.0, 15.0)\,\si{mas}$. The flux ratios with respect to S29 for each of these sources are $0.055$, $0.037$, $0.018,$ and $0.0037$. With $m_\mathrm{K}\left(\mathrm{S29}\right) \simeq 16.6$, this corresponds to a $19.7$, $20.2$, $21.0$ and a $22.7$ magnitude mock source, respectively.

The most important results of the injection tests are summarized in Fig~\ref{fig:results-injection}. From the first row of plots, that is, the injection tests at position 1, close to S300, it is very apparent that the algorithm struggles to properly disentangle both stars. For the highest flux ratio tested, the mock source is found in three runs as a single bright pixel at high significance and in another three smeared out over multiple pixels. Most importantly, however, the scatter around S300 and the amount of flux found in its side lobes increases significantly. The detection of S300 gets cleaner when the flux of the mock source is reduced. But it is already for the second highest flux ratio that the algorithm struggles to robustly detect the mock source and only infers it as flux spread out over multiple pixels at a low significance in six of the ten imaging runs. The result is thus very similar to S300 in the Sgr~A*-centered July images. For a flux ratio of $0.018$, the mock source at the first position is found only once at a low significance. At this point, it is clear that we have exceeded our sensitivity.

At the position of the second mock source, in contrast, the sensitivity is considerably higher. These results are shown in the second row of Fig.~\ref{fig:results-injection}. For a flux ratio of $0.018$ or magnitude of $21.0$, all imaging runs infer flux at the correct position, five of them in a single pixel with high significance and the remaining spread out over multiple neighboring pixels. The faintest magnitude we tested is $22.7$; at this point, only one of the ten imaging runs is able to identify the mock source and we would not be able to claim a detection. Interestingly, however, more flux is attributed to a spurious source in the side lobe west of S55, which is rather close to the injected star. 

\section{Residual images}
\label{appx:residual-images}

\begin{figure*}
\centering
\includegraphics[]{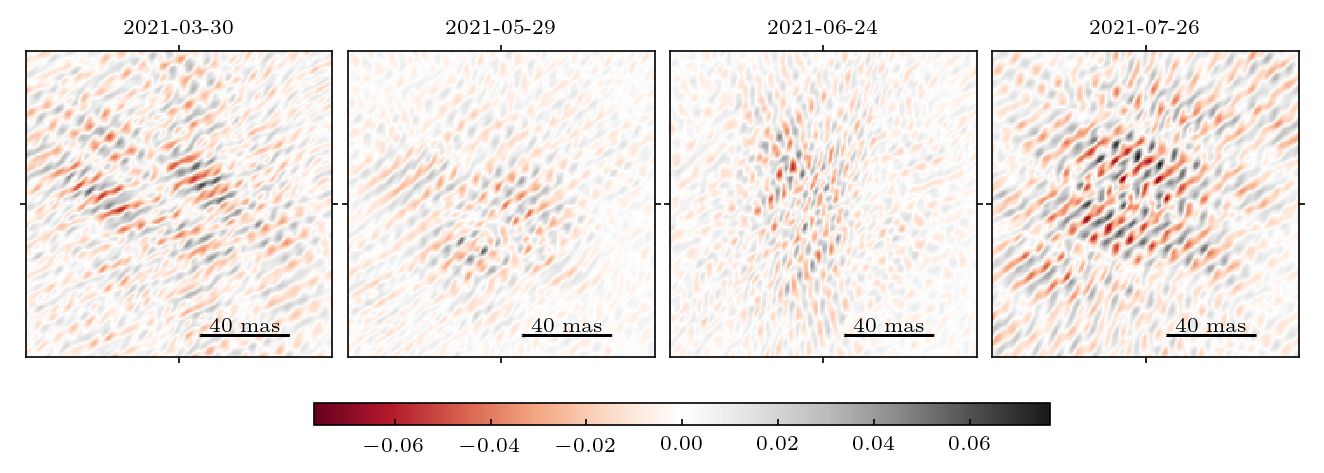}
\caption{Residual images for the four observing epochs in 2021, where north is up, east to the left and the flux is normalized to S29. The images corresponding to the nights depicted here are shown in Fig.~\ref{fig:results-images}. We note that while our posterior model was optimized on closure phases and visibility amplitudes, here we compare it to the full complex visibility.}
\label{fig:appx-residual-images}
\end{figure*}

\begin{table}
\caption{Rms in the central $74 \times 74\,\si{mas}$ of the residual images in Fig.~\ref{fig:appx-residual-images}.}
\label{appx:tab-noise-rms}
\centering
\begin{tabular}{c c c}
Date & rms/S29 & magnitude\\
\hline
2021-03-30 & $1.6\times 10^{-2}$ & $21.1$\\
2021-05-29 & $1.2\times10^{-2}$ & $21.7$ \\
2021-06-24 & $1.3\times 10^{-2}$ & $21.6$\\
2021-07-26 & $2.2\times 10^{-2}$ & $21.0$
\end{tabular}
\tablefoot{We convert the flux ratio with respect to S29 into a K-band magnitude, assuming that $m_\mathrm{K}\left(\mathrm{S29}\right)=16.6$.}
\end{table}

Residual images play a crucial role for the CLEAN algorithm in that they guide the investigator when deciding where to place the clean box. Furthermore, the highest residual inside this box is understood as a physical source, and its signal is subtracted from the visibilities. 

For the \gr-code, residual images do not have the same significance. We worked with closure phases and visibility amplitudes and performed the comparison between the model and data directly in the data space by evaluating the likelihood. Indeed, to be able to compute a residual image, it is necessary to invert the Fourier transform in the response equation, namely, Eq.~(\ref{eq: method-visibilities}), which requires the full phase information of the complex visibilities. Even then, the normalization term, bandwidth smearing and aberration corrections (see Sect.~\ref{sec:method-response} and Appendix~\ref{appx:response}) cannot be accounted for.

To maintain an understanding of the spatial structure of our residuals and to aid in the comparison of other methods, here we provide retroactively computed residual images. These make use of the absolute phase measurement which is provided by GRAVITY but was not considered in the imaging. 

Closure phases are insensitive to a global translation and our model therefore fixes Sgr~A* at the center of the image. In reality, a slight offset from zero is plausible and would reflect itself in the visibility phases. To correct for this offset, we compute a pseudo-dirty image from our model, and fit a Gaussian to its brightest peak. Doing the same in the dirty image, we can align the model with the data by shifting the coordinates of either Gaussian on top of each other.

We also need to fix the normalization term which is implemented in our model. To determine the overall scaling of the residual images, we use the flux in the brightest pixel of the pseudo-dirty image which corresponds to S29. Our final residual images are given by the inversely Fourier transformed residuals, divided by the S29 signal and are shown in Fig.~\ref{fig:appx-residual-images}.

The imaging analysis of GRAVITY GC data with CLEAN in \cite{2021A&A...645A.127G} used the residual images to estimate the noise level as the root mean square (rms) in the central $74 \times 74\,\si{mas}$ of the image. We provide the same values for all images shown in Fig.~\ref{fig:appx-residual-images} in Table~\ref{appx:tab-noise-rms}. We note that all images  are found to be below the noise level reported in \cite{2021A&A...645A.127G}.

Although a comparison with CLEAN based on the residual images is tempting, we caution the reader against overinterpreting these results. Any telescope-based phase errors are reflected in the residual images, even though \gr~is completely insensitive to them. The inversion of the measurement equation to translate the residuals into the image domain cannot account for several important effects such as normalization by the photometric flux, bandwidth smearing, or aberration maps. Finally, the alignment of our model with the absolute position information available from visibility phases is not perfect and may even have the effect of increasing the noise level by itself.

\end{appendix}

\end{document}